\documentclass[a4paper,11pt]{article}
\usepackage{jheppub} 
\usepackage{lineno}

\usepackage{csquotes}
\usepackage{tikz}

\newcommand{\zetab}{\bar{\zeta}}
\newcommand{\dd}{\mathrm{d}}

\newcommand{\Ef}[3]{{\textrm{E}_4}\!\left(\begin{smallmatrix}#1\\#2\end{smallmatrix};#3\right)}
\newcommand{\gamt}[3]{{\widetilde{\Gamma}}\!\left(\begin{smallmatrix}#1\\#2\end{smallmatrix};#3\right)}
\newcommand{\gamtu}[3]{{\widetilde{\underline{\Gamma}}}\!\left(\begin{smallmatrix}#1\\#2\end{smallmatrix};#3\right)}
\newcommand{\gampm}[3]{{\widetilde{\Gamma}}_\pm\!\left(\begin{smallmatrix}#1\\#2\end{smallmatrix};#3\right)}

\title{\boldmath An Elliptic One-Loop Amplitude in Anti-de-Sitter Space}

\author{Sven F. Stawinski}
\affiliation{Bethe Center for Theoretical Physics, Universität Bonn, D-53115, Germany}

\emailAdd{sstawins@uni-bonn.de}

\abstract{We present full analytic results for the four-point one-loop amplitude of a conformally coupled scalar in four-dimensional Anti-de-Sitter space dual to a primary operator with scaling dimension 1. The computation is based on an intriguing recent discovery, connecting Witten diagrams and flat-space Feynman integrals, which led to an expression of the amplitude of interest as a pure combination of single-valued multiple polylogarithms and an integral which cannot be reduced to multiple polylogarithms. We explicitly evaluate that integral in terms of elliptic multiple polylogarithms, finding that it is not manifestly single-valued unlike the polylogarithmic contributions to the amplitude. Further we compute the symbol of the integral and observe similar structures as for (elliptic) flat-space amplitudes. The result presented here adds to the relatively short list of explicitly known position space curved-space amplitudes beyond tree level, and constitutes the first curved-space amplitude evaluated in terms of elliptic multiple polylogarithms.}

\begin{document}
\preprint{ BONN-TH-2023-10 }
\maketitle
\flushbottom

\section{Introduction}
\label{sec:intro}

Over the last couple of decades, there has been a tremendous increase in our understanding of flat-space scattering amplitudes both at tree/integrand level as well as at the level of the full integrated amplitude, see e.g.~ref.~\cite{sagexOverview} for a recent overview of the field (also see Refs.~\cite{elvangHuang,hennPlefkaNewBook} for textbook introductions). The progress in the latter case has largely been driven by our increase in the understanding of Feynman integrals and in particular their connection to certain complex geometries, see ref.~\cite{weinzierlBook} for an extensive and pedagogical overview.\\
The simplest geometry appearing is the unique genus 0 surface, the Riemann sphere, which is the geometry underlying all one-loop Feynman integrals (at least for integer propagator powers). Every Feynman integral falling into this class is naturally evaluated in terms of multiple polylogarithms (MPLs) \cite{goncharovMPLs1,goncharovMPLs2} and our understanding of the algebraic properties of these special functions has greatly benefited our ability to compute Feynman integrals \cite{spradlinVolovichSymbol,claudeCoaction}. Starting at two loops, there are also Feynman integrals connected to surfaces of higher genus as well as higher dimensional spaces, like Calabi-Yau geometries, see ref.~\cite{beyondMPLReview} for a recent review. \\
The arguably simplest geometries beyond the Riemann sphere are the genus 1 surfaces, namely elliptic curves. In the recent years integrals connected to elliptic curves have received a lot of attention and we have gained a much better understanding of these integrals. In particular, a class of special functions, called elliptic multiple polylogarithms (eMPLs), has been defined \cite{levineMPLs1,brownLevineMPLs1,broedelEMZV,claudeElliptic1}, generalizing MPLs to the genus 1 case (see also ref.~\cite{schlottererHigherGenus2} for recent progress at higher genus).\\
It is a natural question to ask if it is possible to carry over the tools developed for flat-space amplitudes to also better understand scattering amplitudes on curved backgrounds. The most natural curved spaces to investigate are the maximally symmetric spaces, namely (Anti-)de-Sitter spaces. These are also otherwise well motivated, as de Sitter space is interesting from a cosmological perspective while Anti-de-Sitter (AdS) space has received a tremendous amount of attention since the discovery of the AdS/CFT correspondence \cite{maldacenaAdSCFT}.\\
However, many things we take for granted in flat space are more subtle in curved spaces, as for example the definition of a scattering amplitude or even the definition of a particle~\cite{birrellDavies}. For example in AdS space there is a timelike boundary, and hence there is no clear notion of in- and out-states which would require infinite separation. The first proposal for the definition of an AdS scattering amplitude was given in Refs.~\cite{polchinskiSMatrixFromAdS,giddingsAdSSmatrix,giddingsAdSSmatrix2}, where it was argued that one should consider correlation functions of the field operators inserted on the boundary. This mimics the infinite separation important for flat-space amplitudes and is natural from an AdS/CFT point of view, as these boundary correlation functions are equal to conformal correlation functions of the dual primary operators in the dual conformal field theory (CFT). We will be referring to this definition when we talk about AdS amplitudes in the rest of this paper. \\
The computation of these boundary correlation functions, however, is notoriously difficult. This is well illustrated by the fact that there lie almost twenty years between the first considerations of four-point correlation functions \cite{freedmanCommentsOn4ptFunctions,dHokerFreedmanzIntegrals} and the full computation of the simplest class of holographic correlation functions, namely those involving 1/2-BPS operators \cite{fullHalfBPSCorrelator}, see also ref.~\cite{sagexHalfBPS} for a review. \\
More recently there has also been a proposal of interpreting the Mellin transform of boundary correlation functions, the so-called Mellin amplitudes, as AdS scattering amplitudes \cite{mackMellin1,mackMellin2,penedonesCorrelatorsasAmplitudes,penedonesNaturalLanguageForCorrelators}. This is motivated by their analytic similarities to flat-space amplitudes and their flat-space limit. \\
The analytic simplicity of Mellin amplitudes allowed more progress \cite{paulosFeynmanRulesMellin,fitzpatrickAnalyticityHolography,volovichFeynmanRulesMellin,penedonesConformalRegge,penedonesMellinFactorization}, which particularly picked up after the development of a powerful bootstrap approach \cite{zhouMellinBootstrap1,zhouMellinBootstrap2,zhouBootstrap20Theory,zhouBootstrap3d,zhouBootstrapEightSupercharges,zhouBootstrapAdS3}. Note that there is also a similar bootstrap approach directly in position space \cite{zhouMellinBootstrap1,zhouMellinBootstrap2,zhouBootstrapAdS3} which however relies on a certain truncation condition on the spectrum \cite{dHokerFreedmanzIntegrals} that is fulfilled in certain supersymmetric theories but cannot be assumed in general.\\
With most of these efforts being focused on tree-level computations there has recently also been some progress on loop corrections for AdS amplitudes corresponding to non-planar $\mathcal{O}(1/N^2)$ corrections in the dual CFT, see ref.~\cite{pierre} and references therein. These however mostly rely on analytic conformal bootstrap methods to perform the computations on the CFT side, see ref.~\cite{analyticCFTBootstrapReview} for a review. On the other hand explicit bulk calculations in position space are quite rare \cite{carmiAdSLoops1,carmiAdSLoops2,sachsAdSLoops1,sachsAdSLoops2,sachsdSLoops,pierre,fichet1,fichet2}. \\
Recently, however, in ref.~\cite{pierre} it was found that in a very specific setup, to be introduced in section \ref{sec:amplitude}, the Witten diagrams needed for the position space calculation, take a form very similar to flat-space Feynman integrals. This allowed the authors to use methods well-known from flat-space multi-loop computations to evaluate the four-point boundary correlation function up to one-loop order for the boundary condition $\Delta=2$ (corresponding to the scaling dimension of the corresponding primary operator in the dual CFT on the boundary). This result could be fully expressed in terms of single-valued multiple polylogarithms \cite{svPolylogs1,zagierSVPolylogs,claudeCleanSVPolylogs}. \\
It was furthermore pointed out that the amplitude is more complicated for $\Delta=1$, since it involves an integral over a square root defining an elliptic curve in the computation of the finite part of the one-loop contribution. The main goal of this paper is to compute this integral in terms of elliptic multiple polylogarithms, allowing us to present full analytic results for the one-loop four-point amplitude for $\Delta=1$. To the best of our knowledge this is the first explicit computation of a curved-space amplitude/correlation function in terms of eMPLs. We find that the result of this integral is no longer manifestly single-valued, in contrast with the polylogarithmic contributions to the amplitude. Furthermore we will compute the symbol of the result and find certain structures that were also observed in flat-space amplitudes.\\
The paper is organised as follows. In section \ref{sec:elliptic} we will very briefly review the mathematics of elliptic curves in order to then introduce elliptic multiple polylogarithms and their symbol. We will then, in section \ref{sec:amplitude}, review the setup and sketch the computation of the one-loop amplitude as given in ref.~\cite{pierre}, resulting in an expression involving an elliptic integral. In section \ref{sec:ellipticIntegral} we will then evaluate this integral in terms of elliptic multiple polylogarithms, which allows us to then present and discuss the full analytic result for the one-loop amplitude. In section \ref{sec:symbol} we will compute and analyze the symbol of the elliptic integral. Finally, in section \ref{sec:conclusions}, we will draw our conclusions and point to possible future directions.

\section{Review of elliptic curves, elliptic multiple polylogarithms and their symbol}
\label{sec:elliptic}

The main result of this paper is an expression for (the finite part of) a one-loop AdS amplitude in terms of elliptic multiple polylogarithms (eMPLs), which are integrals over rational functions on  an elliptic curve and therefore generalize the well-known multiple polylogarithms \cite{goncharovMPLs1,goncharovMPLs2} (MPLs). In this section we will give a brief review of the mathematical basics of elliptic curves and the definition of elliptic multiple polylogarithms and their symbols. 

\subsection{Elliptic curves}
We will start by discussing the minimum of mathematical material necessary in order to introduce eMPLs. For a more complete and rigorous introduction, see e.g.~ref.~\cite{silverman}. We will largely follow the conventions of ref.~\cite{claudeEllipticPure}.\\
For our purposes an elliptic curve can be understood as the zero set of a fourth order polynomial equation of the form\footnote{A polynomial of higher degree would define a so-called hyperelliptic curve, in which there has been an increasing amount of interest in the context of flat-space amplitudes \cite{hyperellipticFI1,hyperellipticFI2,vanhoveGeometryFI,weinzierlGenusDrop,schlottererHigherGenusPolylogs,schlottererHigherGenus2}.}
\begin{equation}
    \mathcal{E}=\left\{ (x,y)\in\mathbb{C}^2~|~y^2=P_4(x) \right\},
\end{equation}
where $P_4(x)$ is a monic polynomial of degree 4 with non-degenerate roots $a_i$. We will refer to the roots as the branch-points of the elliptic curve. The coordinates $(x,y)$ should be understood as an affine chart $(x:y:1)$ in the projective space $\mathbb{CP}^2$.\\
Note that an elliptic curve of this form is always birationally equivalent to a curve defined by a polynomial of third order, which is the case usually studied in mathematics. For example the transformation
\begin{align}
    x&=\frac{a_1(a_2-(a_2-a_4)X)-a_2 a_4}{a_1-a_4-(a_2-a_4)X}, \\
    y&=\frac{(a_1-a_2)(a_1-a_4)(a_2-a_4)\sqrt{(a_1-a_3)(a_2-a_4)}Y}{(a_1-a_4-(a_2-a_4)X)^2},
\end{align}
brings the elliptic curve above into the so-called Legendre form
\begin{equation}
    Y^2=X(X-1)(X-\lambda),
\end{equation}
with 
\begin{equation}
    \lambda=\frac{(a_1-a_4)(a_2-a_3)}{(a_1-a_3)(a_2-a_4)}.
\end{equation}
In the following we will assume that all of the roots $a_i$ are real and we will order them according to $a_1<a_2<a_3<a_4$. This is the case that we will encounter later and we would only have to perform minor changes in the following discussion for other cases, see e.g.~ref.~\cite{claudeEllipticParameterInts}. Let us now motivate and define various quantities related to a fixed elliptic curve.\\
There is a unique (up to a constant factor) holomorphic differential on the curve given by $\mathrm{d}x/y$. Since $y$ is given by a square root of a fourth order polynomial with distinct roots, its Riemann surface is a complex torus. The torus admits two independent cycles and hence two periods, which we will define as follows\footnote{The prefactor is chosen such that the periods have unit normalization in the Legendre coordinates $(X,Y)$.}
\begin{equation}
    \omega_1=2c_4\int_{a_2}^{a_3}\frac{\mathrm{d}x}{y},~\quad\omega_2=2c_4\int_{a_1}^{a_2}\frac{\mathrm{d}x}{y},
\end{equation}
with
\begin{equation}
    c_4=\frac{1}{2}\sqrt{(a_1-a_3)(a_2-a_4)}.
\end{equation}
 The branches of the square root when integrating $y=\sqrt{P_4(x)}$ along the real axis are chosen as follows:
\begin{equation}
    \sqrt{P_4(x)}=\sqrt{|P_4(x)|}\times 
    \begin{cases} 
    -1 & -\infty < x\leq a_1 \\
    -i & a_1 < x\leq a_2 \\ 
    1  & a_2 < x\leq a_3 \\
    i  & a_3 < x\leq a_4 \\
    -1 & a_4 < x\leq \infty
    \end{cases} .
\end{equation}
Let us also quickly point out another important quantity that in fact uniquely characterizes an elliptic curve, the so-called $j$-invariant
\begin{equation}
    j=256\frac{(1-\lambda(1-\lambda))^3}{\lambda^2(1-\lambda)^2}.
\end{equation}
Two given elliptic curves are isomorphic if and only if their $j$-invariants coincide.\\
As will be sketched below, one can show that an elliptic curve with periods $\omega_1,\omega_2$ is equivalent to a complex torus $\mathbb{C}/\Lambda$, where $\Lambda=\omega_1 \mathbb{Z}\oplus \omega_2 \mathbb{Z}$ is a lattice spanned by the periods. It turns out that two elliptic curves are isomorphic if and only if the lattices corresponding to the associated tori are homothetic, i.e.~equal up to a constant factor. Hence, without loss of generality, we can rescale the lattice and consider the torus $\mathbb{C}/\Lambda_\tau$, $\Lambda_\tau=\mathbb{Z}\oplus \tau \mathbb{Z}$, where we defined the modular parameter
\begin{equation}
    \tau=\frac{\omega_2}{\omega_1} \in \mathbb{H},
\end{equation}
taking values on the upper half plane. To see how the identification between an elliptic curve and the torus works, let us define the Weierstraß $\wp$ function
\begin{equation}
    \wp(z;\tau)=\frac{1}{z^2}+\sum_{(m,n)\in\mathbb{Z}^2\setminus\{(0,0)\}}\left( \frac{1}{(z+m+n\tau)^2} - \frac{1}{(m+n\tau)^2}\right).
\end{equation}
The Weierstraß $\wp$ function is an example of an elliptic function (in $z)$, i.e.~a doubly periodic function defined on $\mathbb{C}$. It satisfies an addition theorem \cite{langEllipticCurves}
\begin{equation}
    \wp(z+w;\tau)=\frac{1}{4}\left(\frac{\wp'(z;\tau)-\wp'(w;\tau)}{\wp(z;\tau)-\wp(w;\tau)}\right)^2-\wp(z;\tau)-\wp(w;\tau),
\end{equation}
and the Weierstraß differential equation
\begin{equation}
    \wp'^2=4\wp^3-g_2\wp -g_3=4(\wp-e_1)(\wp-e_2)(\wp-e_3),
\end{equation}
where $g_i,e_i$ are functions of $\tau$ but constant in $z$. \\
It is clear from the differential equation that for any $z\in\mathbb{C}/\Lambda_\tau$ for fixed $\tau$, the point $(x,y)=(\wp(z),\wp'(z))$ lies on the elliptic curve defined by the equation
\begin{equation}
    y^2=4x^3-g_2(\tau)x-g_3(\tau),
\end{equation}
or
\begin{equation}
    y^2=4(x-e_1(\tau))(x-e_2(\tau))(x-e_3(\tau)).
\end{equation}
It follows from the properties of the Weierstraß $\wp$ function that these are genuine elliptic curves, i.e.~the roots are non-degenerate. Conversely it is easy to see that given one of the two points $(x_0,\pm y_0)$ on the elliptic curve, the point
\begin{equation}
    z_\pm=\pm\int_\infty^{x_0}\frac{\mathrm{d}x}{y}\text{ mod }\Lambda,
\end{equation}
satisfies $(\wp(z_\pm;\tau),\wp'(z_\pm;\tau))=(x_0,\pm y_0)$. Thus we have a bijection between the elliptic curve $\mathcal{E}$ and the torus $\mathbb{C}/\Lambda$ given by the maps
\begin{align}
    \mathcal{E}\rightarrow \mathbb{C}/\Lambda,~&(x_0,\pm y_0)\mapsto\pm \int_\infty^{x_0}\frac{\mathrm{d}x}{y}\text{ mod }\Lambda, \\
    \mathbb{C}/\Lambda\rightarrow\mathcal{E},~&z\mapsto (\wp(z),\wp'(z)) .
\end{align}
Using the fact that every elliptic curve is birationally equivalent to one in Weierstraß form we can transform this correspondence to the case of an elliptic curve defined by a quartic polynomial. One finds \cite{claudeElliptic1}
\begin{align}
    \mathcal{E}\rightarrow \mathbb{C}/\Lambda_\tau,~&(x_0,\pm y_0)\mapsto \pm z_{x_0}\equiv\pm\frac{c_4}{\omega_1}\int_{a_1}^{x_0}\frac{\mathrm{d}x}{y}\text{ mod }\Lambda_\tau, \\
    \mathbb{C}/\Lambda_\tau\rightarrow\mathcal{E},~&z\mapsto (\kappa(z;\Vec{a}),c_4\kappa'(z;\Vec{a})) .
    \label{eq:torusCorrespondence}
\end{align}
Here, $\Vec{a}=(a_1,a_2,a_3,a_4)$ denotes the vector of branch points and the function $\kappa$ is defined by
\begin{equation}
    \kappa(z,\Vec{a})=\frac{-3a_1 a_{13} a_{24}\wp(\omega_1 z;\tau)+a_1^2\Bar{s}_1-2a_1\Bar{s}_2+3\Bar{s}_3}{-3 a_{13} a_{24}\wp(\omega_1 z;\tau)+3a_1^2-2a_1\Bar{s}_1+\Bar{s}_2},
\end{equation}
with the abbreviations $a_{ij}=a_i-a_j$ and $\Bar{s}_n=s_n(a_2,a_3,a_4)$ denoting symmetric polynomials of degree $n$.

\subsection{Elliptic multiple polylogarithms}
In the last subsection we introduced the bare minimum of material on elliptic curves that now allows us to introduce elliptic multiple polylogarithms (eMPLs) \cite{levineMPLs1,brownLevineMPLs1,broedelEMZV,claudeElliptic1}. We will largely follow Refs.~\cite{claudeElliptic1,claudeEllipticPure}. \\
Elliptic multiple polylogarithms form a natural function basis for integrals over rational functions on an elliptic curve. They are defined as iterated integrals over kernels $g^{(n)}(z,\tau)$ defined by a generating series, the so-called Eisenstein-Kronecker series
\begin{equation}
    F(z,\alpha,\tau)=\frac{\theta_1 '(0,\tau)\theta_1(z+\alpha,\tau)}{\theta_1(z,\tau)\theta_1(\alpha,\tau)}=\frac{1}{\alpha}\sum_{n=0}^\infty g^{(n)}(z,\tau) \alpha^n .
\end{equation}
Here 
\begin{equation}
    \theta_1(z,\tau)=2 q^{1/8}\sin(\pi z)\prod_{n=1}^\infty (1-q^n)(1-2q^n\cos(2\pi z)+q^{2n}),
\end{equation}
is the odd Jacobi theta function and $q=e^{2\pi i\tau}$. The kernels have definite parity
\begin{equation}
    g^{(n)}(-z,\tau)=(-1)^n g^{(n)}(z,\tau),
\end{equation}
and satisfy the following (quasi-)periodicity properties
\begin{align}
    &g^{(n)}(z+1,\tau)=g^{(n)}(z,\tau),&g^{(n)}(z+\tau,\tau)=\sum_{k=0}^n \frac{(-2\pi i)^k}{k!}g^{(n-k)}(z,\tau) .
\end{align}
Inside the fundamental domain spanned by 1 and $\tau$ only the kernel $g^{(1)}(z,\tau)$ has a simple pole with unit residue at the origin, whereas the kernels $g^{(n)}(z,\tau)$ with $n\geq 2$ are regular. However, also the higher weight kernels acquire poles outside of the fundamental domain e.g~at $z=\tau$ due to quasi-periodicity.\\
Note that, since the kernels do not have full double periodicity, they are strictly speaking not well-defined on the torus, but only on its universal cover, which is $\mathbb{C}$. One can remedy this by adding a factor to the Eisenstein-Kronecker series that cancels the lack of double-periodicity at the cost of introducing an anti-holomorphic dependence \cite{broedelEMZV}. In the following we will however stick to the meromorphic kernels defined above. \\
Elliptic multiple polylogarithms (eMPLs) are now defined as iterated integrals over the kernels we just introduced
\begin{equation}
    \gamt{n_1 & \dots & n_k}{z_1 & \dots & z_k}{z,\tau}=\int_0^z\mathrm{d}z'\,g^{(n_1)}(z'-z_1,\tau) \gamt{n_2 & \dots & n_k}{z_2 & \dots & z_k}{z',\tau},
\end{equation}
with the recursion starting with $\gamt{}{}{z,\tau}=1$. We will refer to $k$ as the length and to $\sum_{i=1}^k n_i$ as the weight of the eMPL. \\
As iterated integrals, eMPLs satisfy some general properties \cite{chenIteratedIntegrals}, the most important for us will be that they satisfy a shuffle algebra
\begin{equation}
    \widetilde{\Gamma}(A_1,\dots,A_k;z,\tau)\widetilde{\Gamma}(A_{k+1},\dots,A_{k+l};z,\tau)=\sum_{\sigma\in\Sigma(k,l)}\widetilde{\Gamma}(A_{\sigma(1)},\dots,A_{\sigma(k+l)};z,\tau),
\end{equation}
with $A_i=\left(  \begin{smallmatrix} n_{i} \\ z_{i} \end{smallmatrix} \right)$ and where $\Sigma(k,l)$ denotes the set of shuffles of the sets $\{1,\dots,k\},\{k+1,\dots,k+l\}$, i.e.~the collection of ordered unions of these two sets that preserve the order in the respective sets.\\
Note that since the kernel $g^{(1)}(z,\tau)$ has a pole at $z=0$, there might be end-point divergences making the definition above ill-defined for $(n_k,z_k)=(1,0)$. We can always use the shuffle algebra to separate divergent eMPLs into finite ones and powers of the basic divergent eMPL which we regularize in the following way
\begin{equation}
    \gamt{1}{0}{z,\tau}=\log\left(1-e^{2\pi iz}\right)-2\pi iz+\int_0^z\mathrm{d}z'\left( g^{(1)}(z',\tau)-\frac{2\pi i}{e^{2\pi i z'}-1} \right)-\log\left(\frac{-2\pi i}{\omega_1} \right).
\end{equation}
Note that this regularization differs from the one chosen in ref.~\cite{claudeEllipticPure} by an additional term which will lead to a nicer form of the result of the integral to be computed in section \ref{sec:ellipticIntegral}.\\
There is also a different definition of elliptic multiple polylogarithms phrased not on the torus but directly on the elliptic curve. They are also given as iterated integrals over kernels taken from an infinite set. In the following we will only need the following kernels of weights 0 and 1:
\begin{align}
    \psi_0(0,x)&=\frac{c_4}{y}, \\
    \psi_1(c,x)&=\frac{1}{x-c}, \\
    \psi_{-1}(c,x)&=\frac{y_c}{y(x-c)}.
\end{align}
Here, $y_c=\sqrt{P_4(c)}$ and $c\in\mathbb{C}$. Expressions for the other kernels can be found in ref.~\cite{claudeElliptic1}. The eMPLs on the elliptic curve are now again defined as iterated integrals over these kernels
\begin{equation}
    \Ef{n_1 & \dots & n_k}{c_1 & \dots & c_k}{x,\Vec{a}}=\int_0^x\mathrm{d}x'\,\psi_{n_1}(c_1,x')\Ef{n_2 & \dots & n_k}{c_2 & \dots & c_k}{x',\vec{a}},
\end{equation}
with $\Ef{}{}{x,\Vec{a}}=1$. It was shown in ref.~\cite{claudeElliptic1} that the space of functions spanned by the two different definitions of eMPLs is the same, i.e.~one can express every eMPL of the one sort as a $\mathbb{C}$-linear combination of eMPLs of the other sort. This is achieved by relating the differential forms that are integrated over using the map between the elliptic curve and the torus and general facts about elliptic functions. We will only need the following relations
\begin{align}
    \mathrm{d}x\,\psi_1(c,x)&=\mathrm{d}z\,\left[ g^{(1)}(z-z_c,\tau)+g^{(1)}(z+z_c,\tau)-g^{(1)}(z-z_*,\tau)-g^{(1)}(z+z_*,\tau) \right], \\
    \mathrm{d}x\,\psi_{-1}(c,x)&=\mathrm{d}z\,\left[ g^{(1)}(z-z_c,\tau)-g^{(1)}(z+z_c,\tau)+g^{(1)}(z_c-z_*,\tau)+g^{(1)}(z_c+z_*,\tau) \right],
\end{align}
where
\begin{equation}
    z_*=\frac{c_4}{\omega_1}\int_{a_1}^\infty\frac{\mathrm{d}x}{y},
\end{equation}
is the torus point corresponding to the point at infinity on the elliptic curve. Note that MPLs are a special case of eMPLs, since we have
\begin{equation}
    G(c_1,\dots\,c_n;z)=\Ef{1 & \dots & 1}{c_1 & \dots & c_n}{z,\Vec{a}},
\end{equation}
which allows us to write certain combinations of eMPLs on the torus as ordinary logarithms.
\subsection{The symbol of elliptic multiple polylogarithms}
\label{ssec:symbolReview}
The symbol of a transcendental function \cite{goncharovSymbol} is a tensor encoding the differential equation of the function and can, roughly speaking, be understood as a decomposition into \enquote{elementary building blocks}, the so-called symbol letters. The first entries of the symbol encode discontinuities of the function, while the last entries encode derivatives. It was first introduced into the physics literature in the context of multiple polylogarithms in ref.~\cite{spradlinVolovichSymbol}, further developed in ref.~\cite{claudeSymbolsPolygons} and connected to a Hopf algebra structure in ref.~\cite{claudeCoaction}. It was extended to eMPLs in ref.~\cite{claudeEllipticSymbol}. We will in large parts follow the conventions of ref.~\cite{wilhelmSymbolPrime}.\\
Generally speaking, if we consider an iterated integral of length $k$ where the total differential consists of one-forms $\omega_i$ and integrals of strictly lower length $I^{(i)}_{k_i},~k_i<k$, schematically written as
\begin{equation}
    \dd I_k=\sum_i I_{k_i}^{(i)} \,\omega_i,
\end{equation}
one can recursively define the symbol as
\begin{equation}
    \mathcal{S}(I_k)=\sum_i  \mathcal{S}(I_{k_i}^{(i)})\otimes \omega_i,~k>0,\text{ and } \mathcal{S}(I_0)=I_0,
\end{equation}
the symbol letters then being the differential one-forms $\omega_i$. For multiple polylogarithms for example the symbol letters are $\dd\!\log$ forms of rational functions of the kinematics
\begin{equation}
    \omega_i=\dd\!\log\mathcal{R}_i.
\end{equation}
To find the symbol of eMPLs we hence need to consider their total differential. It is given in closed form by \cite{claudeEllipticSymbol}
\begin{align*}
    &\dd\gamt{n_1 & \dots & n_k}{z_1 & \dots & z_k}{z,\tau}=\sum_{p=1}^{k-1}(-1)^{n_{p+1}}\gamt{n_1 \dots n_{p-1} & 0 & n_{p+2} \dots n_k}{z_1 \dots z_{p-1} & 0 & z_{p+2} \dots z_k}{z,\tau}\omega^{(n_p+n_{p+1})}(z_{p+1}-z_p,\tau) \\
    &\qquad + \sum_{p=1}^k\sum_{r=0}^{n_p+1}\left[ \begin{pmatrix} n_{p-1}+r-1 \\ n_{p-1}-1 \end{pmatrix} \gamt{n_1 \dots n_{p-2} & n_{p-1}+r & n_{p+1} \dots n_k}{z_1 \dots z_{p-2} & z_{p-1} & z_{p+1} \dots z_k}{z,\tau} \omega^{(n_p-r)}(z_{p-1}-z_p,\tau)  \right. \\
    &\qquad\qquad\qquad \left. -\begin{pmatrix} n_{p+1}+r-1 \\ n_{p+1}-1 \end{pmatrix} \gamt{n_1 \dots n_{p-1} & n_{p+1}+r & n_{p+2} \dots n_k}{z_1 \dots z_{p-1} & z_{p+1} & z_{p+2} \dots z_k}{z,\tau} \omega^{(n_p-r)}(z_{p+1}-z_p,\tau)  \right], \addtocounter{equation}{1}\tag{\theequation}
\end{align*}
where we set $(z_0,z_{k+1})=(z,0),\:(n_0,n_{k+1})=(0,0)$ as well as $\left( \begin{smallmatrix} -1 \\ -1 \end{smallmatrix}\right)=1$. Further we have defined the one-forms
\begin{align}
    \omega^{(n)}(z,\tau)&=\dd z\,g^{(n)}(z,\tau)+\frac{n\,\dd\tau}{2\pi i}g^{(n+1)}(z,\tau), \text{ for } n\geq 0, \\
    \omega^{(-1)}(z,\tau)&=-\frac{\dd\tau}{2\pi i}.
\end{align}
Note that the right hand side of the symbol only involves eMPLs of lower length, i.e.~it satisfies a differential equation without a homogeneous term. Hence we could now immediately associate a symbol using the usual recursive definition as done for example in ref.~\cite{claudeEllipticSymbol}. The symbol letters would then be the one-forms defined above. In analogy with the polylogarithmic case where we usually write the symbol letters as $\dd\!\log$ forms and then drop the $\dd$ operator (and often even the log) we would like to write the forms as the primitive of some function which we would then use as symbol letters. As remarked in ref.~\cite{wilhelmSymbolPrime} this is indeed possible since the one-forms $\omega^{(n)}(z,\tau)$ are exact. Hence we define
\begin{equation}
    \omega^{(n)}(z,\tau)=(2\pi i)^{n-1}\dd\Omega^{(n)}(z,\tau),
\end{equation}
where the conventional power of $2\pi i$ is introduced to ensure that all $\Omega^{(n)}(z,\tau)$ have transcendental weight 1, leading to simpler identities between them. Using these new letters to define the symbol we arrive at the definition \cite{wilhelmSymbolPrime}
\begin{align*}
    &\mathcal{S}\left(\gamtu{n_1\dots n_k}{z_1 \dots z_k}{z}\right)=\sum_{p=1}^{k-1}(-1)^{n_{p+1}}\mathcal{S}\left(\gamtu{n_1 \dots n_{p-1} & 0 & n_{p+2} \dots n_k}{z_1 \dots z_{p-1} & 0 & z_{p+2} \dots z_k}{z}\right)\otimes\Omega^{(n_p+n_{p+1})}(z_{p+1}-z_p) \\
    &\quad + \sum_{p=1}^k\sum_{r=0}^{n_p+1}\left[ \begin{pmatrix} n_{p-1}+r-1 \\ n_{p-1}-1 \end{pmatrix} \mathcal{S}\left(\gamtu{n_1 \dots n_{p-2} & n_{p-1}+r & n_{p+1} \dots n_k}{z_1 \dots z_{p-2} & z_{p-1} & z_{p+1} \dots z_k}{z}\right) \otimes\Omega^{(n_p-r)}(z_{p-1}-z_p)  \right. \\
    &\quad\qquad\qquad \left. -\begin{pmatrix} n_{p+1}+r-1 \\ n_{p+1}-1 \end{pmatrix} \mathcal{S}\left(\gamtu{n_1 \dots n_{p-1} & n_{p+1}+r & n_{p+2} \dots n_k}{z_1 \dots z_{p-1} & z_{p+1} & z_{p+2} \dots z_k}{z}\right) \otimes\Omega^{(n_p-r)}(z_{p+1}-z_p)  \right], \addtocounter{equation}{1}\tag{\theequation}
\end{align*}
with $\gamtu{n_1\dots n_k}{z_1 \dots z_k}{z}=(2\pi i)^{k-\sum_{i=1}^k n_i}\gamt{n_1\dots n_k}{z_1 \dots z_k}{z}$, ensuring that all $\widetilde{\underline{\Gamma}}$ have transcendental weight equal to their length, and suppressing $\tau$ dependencies for brevity\footnote{In the following we will often omit the dependence on $\tau$ without further comment.}. If one ends up with a sum of tensors of different lengths one should project onto the length $k$ component. This projection is left implicit above. We can extend the definition of the symbol to the whole algebra of eMPLs by requiring it to be $\mathbb{C}$-linear and map the product of two eMPLs to the shuffle product of their respective symbols. \\
One can give explicit series representations for the new symbol letters $\Omega^{(n)}(z,\tau)$, see ref.~\cite{wilhelmSymbolPrime}. Note that the $\Omega^{(n)}(z,\tau)$ are only defined up to constants i.e.~closed zero forms. We will in the following identify each $\Omega^{(n)}(z,\tau)$ with its equivalence class modulo addition of zero forms, i.e.~we will understand them as cohomology classes.\\
The symbol letters inherit various properties from the kernels appearing in the one-forms
\begin{align}
    \text{parity: }~~~ \Omega^{(n)}(-z,\tau)&=(-1)^{n+1}\Omega^{(n)}(z,\tau), \\
    \text{periodicity: } \Omega^{(n)}(z+1,\tau)&=\Omega^{(n)}(z,\tau), \\
    \text{quasi-periodicity: } \Omega^{(n)}(z+\tau,\tau)&=\sum_{k=0}^{n+1}\frac{(-1)^k}{k!}\Omega^{(n-k)}(z,\tau) .
\end{align}
Remember that these are identities between the cohomology classes, so they only hold modulo constants.\\
Since the symbol letters are primitives of the one-forms consisting of the kernels they are connected to length 1 eMPLs
\begin{equation}
    (2\pi i)^{1-n}\gamt{n}{z_c}{z,\tau}=\Omega^{(n)}(z-z_c,\tau)-\Omega^{(n)}(-z_c,\tau).
\end{equation}
One might be concerned with $\Omega^{(1)}(0,\tau)$ being divergent, inheriting the divergence (and regularization) from $\gamt{1}{0}{z,\tau}$. However remember that in the symbol we are really interested in 
\begin{equation}
    \dd\Omega^{(1)}(0,\tau)=\frac{\dd\tau}{2\pi i}g^{(2)}(0,\tau),
\end{equation}
which is completely finite. \\
There are further identities beyond the symbol letters besides the trivial ones stated above. The simplest ones (and the only ones that we will use in this paper) express certain combinations of letters as a $\mathrm{d}\!\log$ form. Observe for example
\begin{align*}
    \log\left(\frac{b-c}{a-c}\right)&=\int_a^b\frac{\mathrm{d}\sigma}{\sigma-c}=\gamt{1}{z_c}{z_b}+\gamt{1}{-z_c}{z_b}-\gamt{1}{z_*}{z_b}-\gamt{1}{-z_*}{z_b}-(z_b\rightarrow z_a) \\
    &=\Omega^{(1)}(z_b-z_c)+\Omega^{(1)}(z_b+z_c)-\Omega^{(1)}(z_b-z_*)-\Omega^{(1)}(z_b+z_*)-(z_b\rightarrow z_a). \addtocounter{equation}{1}\tag{\theequation}
\end{align*}
Taking the difference of two such equations with the same $a,b$ but different values of $c$ gives a relation where all dependence on $z_*$ drops out. Degenerations of this identity and further identities building upon Abel's addition theorem and the elliptic Bloch relation were investigated in ref.~\cite{wilhelmSymbolPrime}.

\section{The four-point amplitude of a conformally coupled scalar}
\label{sec:amplitude}
The main goal of this paper is to compute a certain AdS loop amplitude in terms of eMPLs as introduced in the preceding section. The computation builds upon a relation between Witten diagrams and flat-space Feynman integrals first pointed out in ref.~\cite{pierre} in the context of a conformally coupled scalar field in $\mathrm{AdS}_4$. We will first review this setup and the mapping to flat-space Feynman integrals. Building upon this we will then sketch the computation of the four-point amplitude up to one-loop order as performed in ref.~\cite{pierre} to which we refer for more details. This will result in an expression containing single-valued multiple polylogarithms and an elliptic integral which we will then evaluate in terms of eMPLs in the following section.

\subsection{The setup}
Let us first define the theory in which the connection between Witten diagrams and Feynman integrals was found. To this end consider the Poincaré patch of Wick rotated (Euclidean) four-dimensional Anti-de-Sitter space ($\mathrm{AdS}_4$) with coordinates $x^\mu=(z,\mathbf{x}^i)$\footnote{Mathematically speaking this is just the Poincaré upper half-plane model of hyperbolic geometry.}. Here $\mathbf{x}\in\mathbb{R}^3$ parametrizes the conformally flat boundary and $z\in\mathbb{R}_{> 0}$ denotes the radial coordinate. The metric takes the form
\begin{equation}
    \mathrm{d}s^2=\frac{\mathrm{d}z^2+\mathrm{d}\mathbf{x}^2}{a^2z^2},
\end{equation}
where $a$ denotes the AdS radius, which we will set to one in the following.\\
We will now couple a massless scalar to this background geometry in a non-minimal way
\begin{equation}
    S=\int_{\mathrm{AdS_4}}\mathrm{d}^4 x\,\sqrt{g}\left(\frac{1}{2}(\partial\phi)^2+\frac{1}{2}\xi R\phi^2\right),
\end{equation}
such that the equations of motion have conformal symmetry (conformal coupling). This is achieved by setting $\xi=\frac{(D-2)}{4(D-1)}$ in $\mathrm{AdS}_D$ \cite{birrellDavies}. Plugging in the value of the Ricci scalar $R=-D(D-1)$ and $D=4$ generates an effective mass $m^2=-2$. To allow for interesting dynamics we also need to add an interaction term. As will become clear later, the simplest interaction for which the mapping to flat-space Feynman integrals works for the scaling and boundary dimensions of interest, is quartic. Adding this leaves us with the action
\begin{equation}
    S=\int_{\mathrm{AdS_4}}\mathrm{d}^4 x\,\sqrt{g}\left(\frac{1}{2}(\partial\phi)^2+\frac{1}{2}m^2\phi^2+\frac{1}{4!}\lambda\phi^4 \right).
\end{equation}
By inspecting the free field equation one can argue that the field has the asymptotic behaviour $\phi(z,\mathbf{x})\sim z^\Delta \phi_0(\mathbf{x})$ with $\Delta(\Delta-d)=m^2$ where $d=3$ is the boundary dimension \cite{wittenAdS}. This gives us the two choices $\Delta=1,2$ corresponding to different boundary conditions, which we will indicate by a subscript $\phi_\Delta(x)$\footnote{Note that both possibilities are inside the unitarity window $\Delta\geq (d-2)/2$ \cite{wittenUnitarityBound}.}. By the AdS/CFT correspondence there is a dual CFT containing a primary operator $\mathcal{O}_\Delta$ with scaling dimension $\Delta$ corresponding to our scalar field. Hence we will also, somewhat loosely, refer to $\Delta$ as the scaling dimension of $\phi$. \\
One might be worried that since our bulk theory does not include gravity, i.e.~no dynamical metric, there is no stress-tensor in the dual CFT, which does not seem to make sense. However, the theory we are considering can still be interesting from an AdS/CFT point of view, as it can approximate a gravitational theory with a strongly self-interacting scalar $\phi$. Since our focus will be on the bulk computation we will not comment on this further, see e.g.~ref.~\cite{holographyCFTPolchinski} for a more elaborate discussion. \\
We want to compute the four-point amplitude in the theory just introduced, which we take to be defined as the correlation function of the field operators inserted on the boundary $\langle \phi_\Delta(\mathbf{x}_1)\phi_\Delta(\mathbf{x}_2)\phi_\Delta(\mathbf{x}_3)\phi_\Delta(\mathbf{x}_4)\rangle$, as already alluded to in the introduction. This equals the CFT correlation function of the dual operators $\langle \mathcal{O}_{\Delta}(\mathbf{x}_1)\mathcal{O}_{\Delta}(\mathbf{x}_2)\mathcal{O}_{\Delta}(\mathbf{x}_3)\mathcal{O}_{\Delta}(\mathbf{x}_4) \rangle$. We will however compute this entirely from the AdS side, i.e.~we will not make use of any boundary information. Note however that the bulk one-loop computation allows the extraction of quantum corrections to the anomalous dimensions of the CFT, see Refs.~\cite{sachsAdSLoops1,sachsAdSLoops2} for details.\\
The computation of correlation functions works analogously to flat space with Witten diagrams \cite{wittenAdS} acting as the AdS analogues of (position-space) Feynman diagrams. For example the four-point amplitude of the conformally coupled scalar considered here has the diagrammatic expansion\\
\begin{align*}
     \langle \phi_\Delta(x_1)\phi_\Delta(x_2)\phi_\Delta(x_3)\phi_\Delta(x_4) \rangle &= 
     \left(\begin{tikzpicture}[baseline]
        \draw (0,0) circle (1);
        \draw (-0.707107,0.707107) -- (0.707107,0.707107);
        \draw (-0.707107,-0.707107) -- (0.707107,-0.707107);
        \filldraw[black] (-0.707107,0.707107) circle (2pt) node[anchor=east]{$x_2~$};
        \filldraw[black] (0.707107,0.707107) circle (2pt) node[anchor=west]{$~x_3$};
        \filldraw[black] (-0.707107,-0.707107) circle (2pt) node[anchor=east]{$x_1~$};
        \filldraw[black] (0.707107,-0.707107) circle (2pt) node[anchor=west]{$~x_4$};
    \end{tikzpicture}+ \text{perms.}\right) - \lambda
    \begin{tikzpicture}[baseline]
        \draw (0,0) circle (1);
        \draw (-0.707107,0.707107) -- (0.707107,-0.707107);
        \draw (-0.707107,-0.707107) -- (0.707107,0.707107);
        \filldraw[black] (0,0) circle (2pt);
        \filldraw[black] (-0.707107,0.707107) circle (2pt) node[anchor=east]{$x_2~$};
        \filldraw[black] (0.707107,0.707107) circle (2pt) node[anchor=west]{$~x_3$};
        \filldraw[black] (-0.707107,-0.707107) circle (2pt) node[anchor=east]{$x_1~$};
        \filldraw[black] (0.707107,-0.707107) circle (2pt) node[anchor=west]{$~x_4$};
    \end{tikzpicture}  \\
    & +\frac{\lambda^2}{2}\left(\begin{tikzpicture}[baseline]
        \draw (0,0) circle (1);
        \filldraw[black] (-0.707107,0.707107) circle (2pt) node[anchor=east]{$x_2~$};
        \filldraw[black] (0.707107,0.707107) circle (2pt) node[anchor=west]{$~x_3$};
        \filldraw[black] (-0.707107,-0.707107) circle (2pt) node[anchor=east]{$x_1~$};
        \filldraw[black] (0.707107,-0.707107) circle (2pt) node[anchor=west]{$~x_4$};
        \filldraw[black] (-0.33,0) circle (2pt);
        \filldraw[black] (0.33,0) circle (2pt);
        \draw (0,0) circle (0.33);
        \draw (-0.707107,0.707107) -- (-0.33,0);
        \draw (-0.707107,-0.707107) -- (-0.33,0);
        \draw (0.707107,0.707107) -- (0.33,0);
        \draw (0.707107,-0.707107) -- (0.33,0);
    \end{tikzpicture} + \text{perms.} \right) +\mathcal{O}(\lambda^3).
    \addtocounter{equation}{1}\tag{\theequation}
\end{align*}
The Feynman rules for these diagrams are almost the same as in flat-space, the only difference being the different form of the propagators. In particular we have to distinguish between propagators connecting a bulk and a boundary point and propagators connecting two bulk points. The (scalar) bulk-to-bulk propagator is given by \cite{dhokerReview}
\begin{equation}
    G_\Delta(x_1,x_2)=\mathcal{N}_\Delta \xi^\Delta {}_2 F_1\left(\frac{\Delta}{2},\frac{\Delta+1}{2};\Delta-\frac{d}{2}+1;\xi^2\right),
\end{equation}
where 
\begin{equation}
    \xi=\frac{2 z_1 z_2}{z_1^2+z_2^2+(\mathbf{x}_1-\mathbf{x}_2)^2},
\end{equation}
and the normalization factor is given by
\begin{equation}
    \mathcal{N}_\Delta=\frac{2^{-\Delta}}{2\Delta-d}\frac{\Gamma(\Delta)}{\pi^{d/2}\Gamma(\Delta-d/2)}.
\end{equation}
We can obtain the bulk-to-boundary propagator by taking the limit $z_2\rightarrow 0$\footnote{Note that the conventions for the normalization vary in the literature, we will stick to the one used in ref.~\cite{pierre}.}
\begin{equation}
    K_\Delta(x_1,\mathbf{x}_2)=\lim_{z_2\rightarrow 0} z_2^{-\Delta}G_\Delta(x_1,x_2)=2^\Delta\mathcal{N}_\Delta \left(\frac{z_1}{z_1^2+(\mathbf{x}_1-\mathbf{x}_2)^2}\right)^\Delta .
\end{equation}
The key observation in ref. \cite{pierre} was that for the setup given above the form of the bulk-to-bulk propagator simplifies and one can express them in terms of the so-called conformal flat-space propagator
\begin{equation}
    G_\mathrm{c}(x_1,x_2)=\frac{z_1z_2}{(x_1-x_2)^2},
\end{equation}
which has a similar form as the usual Feynman propagators. Note that $(x_1-x_2)^2=(z_1-z_2)^2+(\mathbf{x}_1-\mathbf{x}_2)^2$ denotes the Euclidean squared distance and not the contraction with the AdS metric. Plugging in the boundary dimension and scaling dimensions one explicitly finds
\begin{align}
    G_{\Delta=1}(x_1,x_2)&=\frac{1}{(2\pi)^2}\left[G_\mathrm{c}(x_1,x_2)-G_\mathrm{c}(x_1,\sigma(x_2))\right],\\
    G_{\Delta=2}(x_1,x_2)&=\frac{1}{(2\pi)^2}\left[G_\mathrm{c}(x_1,x_2)+G_\mathrm{c}(x_1,\sigma(x_2))\right],
\end{align}
where $\sigma(z,\mathbf{x})=(-z,\mathbf{x})$ is the antipodal map. \\
After expressing all bulk-to-bulk propagators in this form the integrand of a Witten diagram already closely resembles the integrand of a flat-space Feynman integral, with the factors of the radial coordinate $z$ possibly being interpreted as \enquote{linear propagators} $u\cdot x$ with $u=(1,\mathbf{0})$ perpendicular to the boundary. Such linear propagators are also known to show up in flat-space computations, as for example in heavy quark effective theory \cite{neubertHQET} or region expansions of Feynman integrals \cite{smirnovRegionExpansion}, see also ref.~\cite{linearPropsAMFlow}. \\
One might still be worried about the appearance of the antipodal map and the integration running over $\mathbb{R}^3\times\mathbb{R}_{>0}$ rather than $\mathbb{R}^4$. Due to the quartic interaction, however, the integrand will always be even under the action of the antipodal map, and the integration region can thus easily be extended. The two terms differing by the action of the antipodal map then simply add up, see ref.~\cite{pierre} for more details. This then lets the integral properly take the form of a flat-space Feynman integral (possibly with linear propagators).\\
Another issue that can show up is that the integrals might be divergent. This will in particular be the case for the integral considered in this paper. We will employ dimensional regularization to regulate the divergence. This choice however breaks the AdS-symmetry and there are alternative regularization procedures that preserve the AdS isometries, for more details we refer to ref.~\cite{pierre}.\\
In ref.~\cite{pierre} this connection between Witten diagrams and Feynman integrals alluded to above has been used to fully evaluate the four-point amplitude for $\Delta=2$ up to one-loop order in terms of single-valued combinations of multiple polylogarithms. In the following subsection we will consider the amplitude for $\Delta=1$ which is technically more challenging due to the appearance of an elliptic integral. This will then be evaluated in the following section.
\subsection{The four-point amplitude for $\Delta=1$}
As the amplitude for $\Delta=2$ has already been fully computed \cite{pierre}, let us now focus on the amplitude for $\Delta=1$. We will continue to follow ref.~\cite{pierre}. The order $\lambda^0$ part contains no interaction and is thus trivial. Also the order $\lambda^1$ part is easily computed as was explicitly done in ref.~\cite{pierre}. For contact diagrams the connection between Witten diagrams and flat-space Feynman integrals however is even more immediate and very general, as pointed out in ref.~\cite{zhouYangianCorrelators}. In particular the contact diagram here is (up to a prefactor) equal to the box integral in four-dimensional flat space which is well known to be given by the Bloch-Wigner dilogarithm \cite{davydychevBox}. \\
The first non-trivial piece is thus the one-loop contribution given by the three bubble diagrams\\
\begin{align*}
    &\left.\frac{2}{\lambda^2}\langle \phi_\Delta(x_1)\phi_\Delta(x_2)\phi_\Delta(x_3)\phi_\Delta(x_4) \rangle\right|_{\mathcal{O}(\lambda^2),\Delta=1}= \\
    &\qquad\begin{tikzpicture}[baseline]
        \draw (-4,0) circle (1);
        \filldraw[black] (-0.707107-4,0.707107) circle (2pt) node[anchor=east]{$x_2~$};
        \filldraw[black] (0.707107-4,0.707107) circle (2pt) node[anchor=west]{$~x_3$};
        \filldraw[black] (-0.707107-4,-0.707107) circle (2pt) node[anchor=east]{$x_1~$};
        \filldraw[black] (0.707107-4,-0.707107) circle (2pt) node[anchor=west]{$~x_4$};
        \filldraw[black] (-0.33-4,0) circle (2pt);
        \filldraw[black] (0.33-4,0) circle (2pt);
        \draw (0-4,0) circle (0.33);
        \draw (-0.707107-4,0.707107) -- (-0.33-4,0);
        \draw (-0.707107-4,-0.707107) -- (-0.33-4,0);
        \draw (0.707107-4,0.707107) -- (0.33-4,0);
        \draw (0.707107-4,-0.707107) -- (0.33-4,0);
        \node[] at (0-4,-1.5) {$s$-channel};
        \node[] at (0-2,0) {$+$};
        \draw (0,0) circle (1);
        \filldraw[black] (-0.707107,0.707107) circle (2pt) node[anchor=east]{$x_3~$};
        \filldraw[black] (0.707107,0.707107) circle (2pt) node[anchor=west]{$~x_2$};
        \filldraw[black] (-0.707107,-0.707107) circle (2pt) node[anchor=east]{$x_4~$};
        \filldraw[black] (0.707107,-0.707107) circle (2pt) node[anchor=west]{$~x_1$};
        \filldraw[black] (-0.33,0) circle (2pt);
        \filldraw[black] (0.33,0) circle (2pt);
        \draw (0,0) circle (0.33);
        \draw (-0.707107,0.707107) -- (-0.33,0);
        \draw (-0.707107,-0.707107) -- (-0.33,0);
        \draw (0.707107,0.707107) -- (0.33,0);
        \draw (0.707107,-0.707107) -- (0.33,0);
        \node[] at (0,-1.5) {$t$-channel};
        \node[] at (0+2,0) {$+$};
        \draw (+4,0) circle (1);
        \filldraw[black] (-0.707107+4,0.707107) circle (2pt) node[anchor=east]{$x_1~$};
        \filldraw[black] (0.707107+4,0.707107) circle (2pt) node[anchor=west]{$~x_2$};
        \filldraw[black] (-0.707107+4,-0.707107) circle (2pt) node[anchor=east]{$x_3~$};
        \filldraw[black] (0.707107+4,-0.707107) circle (2pt) node[anchor=west]{$~x_4$};
        \filldraw[black] (-0.33+4,0) circle (2pt);
        \filldraw[black] (0.33+4,0) circle (2pt);
        \draw (0+4,0) circle (0.33);
        \draw (-0.707107+4,0.707107) -- (-0.33+4,0);
        \draw (-0.707107+4,-0.707107) -- (-0.33+4,0);
        \draw (0.707107+4,0.707107) -- (0.33+4,0);
        \draw (0.707107+4,-0.707107) -- (0.33+4,0);
        \node[] at (0+4,-1.5) {$u$-channel}; \addtocounter{equation}{1}\tag{\theequation}
    \end{tikzpicture}.
\end{align*}\\
Let us sketch the computation for the $s$-channel diagram. Using the Feynman rules for Witten diagrams we find
\begin{align*}
    W^{\Delta=1}_{\text{bubble},s}&=\frac{4}{(2\pi)^{12}}\int_{\mathbb{R}^4}\frac{\mathrm{d}^D y_1}{(u\cdot y_1)^4}\frac{\mathrm{d}^D y_2}{(u\cdot y_2)^4}\left[ \left(\frac{(u\cdot y_1)(u\cdot y_2)}{(y_1-y_2)^2}\right)^2+\frac{1}{2}\frac{(u\cdot y_1)(u\cdot y_2)}{(y_1-y_2)^2}\right] \\
    &\qquad\times\frac{u\cdot y_1}{(y_1-\mathbf{x}_1)^2}\frac{u\cdot y_1}{(y_1-\mathbf{x}_2)^2}\frac{u\cdot y_2}{(y_2-\mathbf{x}_3)^2}\frac{u\cdot y_2}{(y_2-\mathbf{x}_3)^2}, \addtocounter{equation}{1}\tag{\theequation}
\end{align*}
where the prefactor comes from the normalization of the propagators, the factor from extending the integration to all of $\mathbb{R}^4$ and a symmetry factor of 1/2\footnote{Note that we have added the symmetry factor compared to e.g.~eq.~(4.13) in ref.~\cite{pierre}. This leads to the overall factor in the previous and also in later equations differing by a factor of 2 from the one in the corresponding equations in ref.~\cite{pierre}. This will not further be pointed out.}. Further we set $D=4-2\epsilon$ to regulate divergences.
One can now use the conformal symmetry in the boundary coordinates to express the integral in terms of conformal invariants, for the details see ref.~\cite{pierre}. One gets
\begin{align*}
    W^{\Delta=1}_{\text{bubble},s}&=\frac{4}{(2\pi)^{12}}\frac{\zeta\zetab}{x_{12}^2 x_{34}^2}\int_{\mathbb{R}^4}\frac{\mathrm{d}^D y_1}{(u\cdot y_1)^4}\int_{\mathbb{R}^4}\frac{\mathrm{d}^D y_2}{(u\cdot y_2)^4}\left[ \left(\frac{(u\cdot y_1)(u\cdot y_2)}{(y_1-y_2)^2}\right)^2+\frac{1}{2}\frac{(u\cdot y_1)(u\cdot y_2)}{(y_1-y_2)^2}\right] \\
    &\qquad\times\left( \frac{1}{(y_1-u_1)^2(y_2-u_1)^2} \right)^{-2\epsilon} \frac{(u\cdot y_1)^2(u\cdot y_2)^2}{y_1^2(y_1-u_\zeta)^2(y_2-u_1)^2} \addtocounter{equation}{1}\tag{\theequation},
\end{align*}
with the vectors 
\begin{equation}
    u_1=(0,1,0,0),\qquad u_\zeta=\left(0,\frac{\zeta+\Bar{\zeta}}{2},\frac{\zeta-\Bar{\zeta}}{2i},0\right),
\end{equation} 
where $\zeta,\Bar{\zeta}$ are connected to the conformal cross ratios
\begin{equation}
    u=\frac{x_{12}^2 x_{34}^2}{x_{14}^2 x_{23}^2}=\zeta\Bar{\zeta},\qquad v=\frac{x_{13}^2 x_{24}^2}{x_{14}^2 x_{23}^2}=(1-\zeta)(1-\Bar{\zeta}),
    \label{eq:crossRatios}
\end{equation}
and $x_{ij}^2=(\mathbf{x}_i-\mathbf{x}_j)^2$. By expanding the square brackets above we find two terms, the first of which is divergent and was already computed in ref.~\cite{pierre} by directly integrating the Feynman parameter representation yielding
\begin{align*}
    W^{\Delta=1}_{\text{bubble},s,\text{div}}=-\frac{\pi^{-2\epsilon}e^{-2\gamma_\mathrm{E}\epsilon}}{(2\pi)^{8}}\frac{\zeta\Bar{\zeta}}{4x_{12}^2 x_{34}^2}&\left( \frac{1}{\epsilon}\frac{2\mathrm{BW}(\zeta,\zetab)}{\zeta-\Bar{\zeta}}+\frac{f_1(\zeta,\Bar{\zeta})}{\zeta-\Bar{\zeta}} - \frac{\mathrm{BW}(\zeta,\zetab)}{\zeta-\Bar{\zeta}}\log(\zeta\Bar{\zeta})\right.\\
    &\left.\qquad+\frac{2\mathrm{BW}(\zeta,\zetab)}{\zeta-\Bar{\zeta}}\log[(1-\zeta)(1-\Bar{\zeta})]\right),
    \addtocounter{equation}{1}\tag{\theequation}
\end{align*}
where $\mathrm{BW}(\zeta,\Bar{\zeta})$ is the Bloch-Wigner dilogarithm 
\begin{equation}
    \mathrm{BW}(\zeta,\zetab)=\mathrm{Li}_2(\zeta)-\mathrm{Li}_2(\zetab)+\frac{1}{2}\log(\zeta\zetab)\log\left(\frac{1-\zeta}{1-\zetab} \right),
    \label{eq:blochWigner}
\end{equation}
and $f_1(\zeta,\Bar{\zeta})$ is a pure \cite{nimaPure,hennCanonicalBasis} single valued combination of multiple polylogarithms of weight 3. An explicit expression can be found in ref.~\cite{pierre} or in eq.~\eqref{eq:f1Function}. \\
The divergence is of UV origin and is proportional to the contact diagram. Hence it can be removed by a counter term which renormalizes the coupling $\lambda$, leading to a scale dependence of the coupling. The corresponding beta function was computed in ref.~\cite{pierre} and shown to be consistent with the well-known result in flat-space $\phi^4$ theory.\\
The finite part
\begin{equation}
    W^{\Delta=1}_{\text{bubble},s,\text{fin}}=\frac{2}{(2\pi)^{12}}\frac{\zeta\zetab}{x_{12}^2 x_{34}^2}\int_{\mathbb{R}^4}\frac{\mathrm{d}^4 y_1}{u\cdot y_1}\int_{\mathbb{R}^4}\frac{\mathrm{d}^4 y_2}{u\cdot y_2}\,\frac{1}{(y_1-y_2)^2y_1^2(y_1-u_\zeta)^2(y_2-u_1)^2},
\end{equation}
is more challenging to evaluate. By Feynman parametrizing, changing variables and performing some of the integrations one can bring this into the form \cite{pierre}
\begin{equation}
    W^{\Delta=1}_{\text{bubble},s,\text{fin}}=\frac{1}{(2\pi)^{8}}\frac{u}{2x_{12}^2 x_{34}^2} I(u,v,1),
\end{equation}
where 
\begin{equation}
    I(x,y,z)=\frac{1}{4}\int_0^1\mathrm{d}r\int_0^1\mathrm{d}\sigma\,\frac{B(\sigma)}{(\sigma^2-1)xr^2+((-x+y-z)\sigma^2+x)r+z\sigma^2},
    \label{eq:integralDef}
\end{equation}
and
\begin{equation}
    B(\sigma)=2\mathrm{Li}_2(\sigma)-2\mathrm{Li}_2(-\sigma)+2\log(\sigma)[\log(1-\sigma)-\log(1+\sigma)].
\end{equation}
The steps in the other channels are similar and one ends up with \cite{pierre}
\begin{align*}
    &\left.\frac{4}{\lambda^2}\langle \phi_\Delta(x_1)\phi_\Delta(x_2)\phi_\Delta(x_3)\phi_\Delta(x_4) \rangle\right|_{\mathcal{O}(\lambda^2),\Delta=1}\\
    &\quad=\frac{1}{(2\pi)^{8}}\left( -\frac{1}{\epsilon}\frac{3}{\pi^2}\mathcal{W}_0^{1,4-4\epsilon}(u,v)+\frac{u}{2x_{12}^2x_{34}^2}\sum_{i\in\{s,t,u\}}L_0^{1,i}(u,v)\right. \\
    &\qquad\qquad\qquad\qquad\qquad\left.+\frac{u}{x_{12}^2x_{34}^2}\sum_{i\in\{s,t,u\}}L_0'^{i}(u,v)+\mathcal{O}(\epsilon)\right). \addtocounter{equation}{1}\tag{\theequation}
\end{align*}
The expressions for $\mathcal{W}_0^{1,4-4\epsilon}(u,v)$ and $L_0^{1,i}(u,v)$ can be found in ref.~\cite{pierre}. The important point for us is that they can be written as pure combinations of single-valued MPLs. On the other hand
\begin{equation}
    L_0'^{i}(u,v)=
    \begin{cases}
        I(u,v,1) & i=s, \\
        I(v,1,u) & i=t, \\
        I(1,u,v) & i=u,
    \end{cases}
    \label{eq:integralChannels}
\end{equation}
where $I(x,y,z)$ is the integral defined in eq.~\eqref{eq:integralDef}. Thus the only part of the computation of the four-point amplitude left to perform is the integral $I(x,y,z)$. This integral turns out to involve an elliptic curve and we will evaluate it explicitly in terms of elliptic multiple polylogarithms in the next section.

\section{The elliptic integral and the full one-loop result}
\label{sec:ellipticIntegral}
The goal of this section and main result of this paper is the computation of the elliptic integral found in the last section. After finding an explicit expression in terms of elliptic multiple polylogarithms we are then able to present the full one-loop result for the amplitude of a conformally coupled scalar in $\mathrm{AdS}_4$ as introduced in the last section.

\subsection{The elliptic integral}
To see the appearance of the elliptic curve in the integral \eqref{eq:integralDef} one can solve the $r$-integral to find
\begin{equation}
    I(x,y,z)=-\frac{1}{4\sqrt{\lambda(x,y,z)}}\int_0^1\frac{\mathrm{d}\sigma}{w}B(\sigma)\log\mathcal{R}(\sigma,w).
\end{equation}
Here $\lambda(x,y,z)=x^2+y^2+z^2-2xy-2xz-2yz$ is the Källén function, $\mathcal{R}(\sigma,w)$ is a rational function and $w$ satisfies
\begin{equation}
    w^2=P_4(\sigma)=\sigma^4+\frac{2x(y+z-x)}{\lambda(x,y,z)}\sigma^2+\frac{x^2}{\lambda(x,y,z)},
\end{equation}
and hence defines an elliptic curve with branch points
\begin{equation}
\begin{split}
    a_1&=-\sqrt{\frac{x}{x-\left(\sqrt{y}+\sqrt{z}\right)^2}},\qquad a_3=\sqrt{\frac{x}{x-\left(\sqrt{y}-\sqrt{z}\right)^2}}, \\
    a_2&=-\sqrt{\frac{x}{x-\left(\sqrt{y}-\sqrt{z}\right)^2}},\qquad a_4=\sqrt{\frac{x}{x-\left(\sqrt{y}+\sqrt{z}\right)^2}}. 
    \label{eq:branchPoints}
\end{split}
\end{equation}
One can easily check that the branch points are all real for 
\begin{equation}
    \sqrt{y}+\sqrt{z}<\sqrt{x},\quad y,z>0.
\end{equation}
We will in the following restrict to this region of parameter space such that we can directly follow the conventions laid out in section \ref{sec:elliptic}. Note that this region includes the region relevant for the $u$-channel diagram with kinematics satisfying $\sqrt{u}+\sqrt{v}<1,\,u,v>0$. With minor changes one can also solve the integral in the parameter regions relevant for the $s$- and $t$-channel diagrams with the kinematics satisfying the same constraint. This will be outlined in appendix \ref{sec:stChannel}. Other kinematic regions can be solved similarly. \\
Since we are focusing on the $u$-channel parameter region we can set $x=1$ in the following. To evaluate this integral in terms of eMPLs note that $B(\sigma)$ is a combination of classical polylogarithms and hence can immediately be rewritten as a combination of eMPLs. This can also be achieved for the logarithm. By differentiating we find that
\begin{equation}
    \frac{\mathrm{d}}{\mathrm{d}\sigma}\log\mathcal{R}(\sigma,w)=4\psi_{-1}(0,\sigma).
\end{equation}
In order to integrate this back up again we add and subtract the pole at $\sigma=0$ and then regularize the divergent integral using the usual tangential base-point regularisation (as described e.g.~in ref.~\cite{claudeElliptic1}) to find
\begin{equation}
    \log\mathcal{R}(\sigma,w)=4\int_0^\sigma\mathrm{d}\sigma'[\psi_{-1}(0,\sigma')-\psi_1(0,\sigma')]+4\log\sigma+\log(yz).
\end{equation}
Expressing the logarithm as an integral $\int_1^\sigma\mathrm{d}\sigma'\,\psi_1(0,\sigma')$ and changing variables to the torus, we thus find the desired expression for the logarithm in terms of eMPLs.\\
We would now like to use the shuffle algebra to express the integrand as a linear combination of eMPLs (with each term only containing one eMPL) and using $\mathrm{d}\sigma/w=\mathrm{d}z\,\omega_1/c_4$ to perform the final integral by definition of eMPLs. However, because we introduced 1 as a base point above, we will encounter eMPLs both of the form $\widetilde{\Gamma}(\dots;z-z_0,\tau)$ and $\widetilde{\Gamma}(\dots;z-z_1,\tau)$, which prevents us from directly using the shuffle algebra. However we can make use of the path decomposition formula for iterated integrals \cite{chenIteratedIntegrals} to write the eMPLs of the latter form in terms of eMPLs of the former form and ones of the form  $\widetilde{\Gamma}(\dots;z_1-z_0,\tau)$. These eMPLs now do not depend on the integration variable and can be taken out of the integral. For the other eMPLs we can now make use of the shuffle algebra and perform the final integration. Since the upper limit of the $\sigma$ integration was 1, the eMPLs of the result now all have $z_1-z_0$ as argument and we can nicely use the shuffle algebra to combine them into a linear combination of eMPLs, with each term only containing one.\\
At this point we note that the expression for the integral contains eMPLs with dependence on 4 punctures on the torus $z_{-1},z_0,z_1,z_*$ and on the modular parameter $\tau$. Since the original integral only had two degrees of freedom, however, we expect there to be relations between the punctures and the modular parameter. Indeed one can find numerically (and it is also not hard to prove analytically) that
\begin{equation}
    z_*=\frac{1}{4},\qquad z_{-1}=\frac{1}{4}+\frac{\tau}{2}-Z,\qquad z_0=\frac{1}{4}+\frac{\tau}{2},\qquad z_1=\frac{1}{4}+\frac{\tau}{2}+Z.
    \label{eq:punctures}
\end{equation}
The two degrees of freedom are now the modular parameter $\tau$ and one puncture on the torus defined by
\begin{equation}
    Z=z_1-z_0=\frac{c_4}{\omega_1}\int_0^1\frac{\mathrm{d}\sigma}{w}.
\end{equation}
One might ask to which point on the elliptic curve this belongs, i.e.~what $\kappa(Z)$ is. Using the addition theorem for $\wp$ one can show that
\begin{equation}
    \kappa(Z)=-a_3\sqrt{\frac{a_4^2-1}{a_3^2-1}}.
    \label{eq:kappaOfZ}
\end{equation}
After plugging in the values \eqref{eq:punctures} for the punctures we would now like to bring the result into a form as canonical as possible. By differentiating, using the properties of the kernels and integrating back up one can bring the expression into a form where all eMPLs are of the form $\gamt{n_1 & \dots & n_k}{z_1 & \dots & z_k}{Z}$ with
\begin{equation}
    n_i\in\{0,1\}, \qquad z_i\in\left\{0,\frac{1}{2},\frac{\tau}{2},\frac{1}{2}+\frac{\tau}{2}\right\}.
\end{equation}
The four choices of $z_i$ correspond to precisely the branch points on the elliptic curve. Note that besides the trivial parity and (quasi-)periodicity properties of the kernels we also needed the following non-trivial relation that reduces a kernel with double argument to a sum of kernels with shifted argument \cite{claudePrivate}
\begin{align*}
    g^{(n)}(2z,\tau)&=\frac{2^n}{4}\left[  g^{(n)}\left(z,\tau\right)+ g^{(n)}\left(z+\frac{1}{2},\tau\right)+g^{(n)}\left(z+\frac{\tau}{2}\right) + g^{(n)}\left(z+\frac{1}{2}+\frac{\tau}{2},\tau\right) \right] \\
    &\qquad +\sum_{m=0}^{n-1}\frac{(i\pi)^{n-m}2^{n-2}}{(n-m)!}\left[ g^{(m)}\left(z+\frac{\tau}{2},\tau\right) + g^{(m)}\left(z+\frac{1}{2}+\frac{\tau}{2},\tau\right)  \right].
    \addtocounter{equation}{1}\tag{\theequation}
\end{align*}
From the correspondence \eqref{eq:torusCorrespondence} it is clear that the sign of the torus punctures corresponds to the sign of $y$, i.e.~the branch of the square root. Since we however expect the physical result to not depend on this choice, the result should not depend on the sign of the torus punctures. We can make this manifest by rewriting the result in a \enquote{parity eigenbasis} defined analogously to ordinary eMPLs by
\begin{equation}
    \gampm{n_1\dots n_k}{z_1\dots z_k}{z,\tau}=\int_0^z\mathrm{d}z'\, g^{(n_1)}_\pm(z',z_1,\tau)\gampm{n_2\dots n_k}{z_2\dots z_k}{z',\tau},
\end{equation}
with $\gampm{}{}{z;\tau}=1$ and where now $n_i\in\mathbb{Z}$ can take negative values. The kernels are defined by
\begin{equation}
\begin{split}
    g_\pm^{(0)}(z_1,z_2,\tau)&=1, \\
    g_\pm^{(n)}(z_1,z_2,\tau)&=g^{(|n|)}(z_1-z_2,\tau)+\mathrm{sign}(n)g^{(|n|)}(z_1+z_2,\tau),~n\neq 0.
\end{split}
\end{equation}
Performing this change of function basis one indeed observes that all parity eMPLs with negative $n_i$ drop out which reflects the fact alluded to above that the result is independent of the signs of the points on the torus. \\
We can now realise that only certain combinations of the kernels show up in the eMPLs. Hence we define
\begin{align*}
    \mathrm{d}z\,g^{(0)}_0(z,\tau)&\equiv \mathrm{d}z\, g^{(0)}_\pm(z,0,\tau)= \mathrm{d}z\,  \addtocounter{equation}{1}\tag{\theequation}\equiv \omega_0,\\
    \mathrm{d}z\,g^{(1)}_A(z,\tau)&\equiv \mathrm{d}z\,\left[g^{(1)}_\pm(z,0,\tau)-g^{(1)}_\pm(z,1/2,\tau)\right] =\mathrm{d}\!\log\left( \frac{\sigma+a_4}{\sigma-a_4} \right) \equiv \omega_A ,\addtocounter{equation}{1}\tag{\theequation}\\
    \mathrm{d}z\,g^{(1)}_B(z,\tau)&\equiv \mathrm{d}z\,\left[ g^{(1)}_\pm(z,\tau/2,\tau)-g^{(1)}_\pm(z,1/2+\tau/2,\tau)\right] =\mathrm{d}\!\log\left( \frac{\sigma+a_3}{\sigma-a_3} \right)\equiv \omega_B, \addtocounter{equation}{1}\tag{\theequation}\\
    \mathrm{d}z\,g^{(1)}_C(z,\tau)&\equiv \mathrm{d}z\,\left[g^{(1)}_\pm(z,0,\tau)+g^{(1)}_\pm(z,1/2,\tau)-g^{(1)}_\pm(z,\tau/2,\tau)-g^{(1)}_\pm(z,1/2+\tau/2,\tau) \right] \\
    &=\mathrm{d}\!\log\left( \frac{\sigma^2-a_4^2}{\sigma^2-a_3^2} \right)\equiv \omega_C , \addtocounter{equation}{1}\tag{\theequation}
\end{align*}
where we used the map from the torus to the elliptic curve to write the weight one combinations as $\mathrm{d}\!\log$ forms. \\
Rewriting the result in terms of the new kernels and rewriting the length 1 eMPLs as logarithms using the $\mathrm{d}\!\log$ representation of the kernels, the expression for the integral takes a very compact form (the technical details of this rewriting are deferred to appendix \ref{sec:detailsOfCalculation})
\begin{align*}
    I&=\frac{\omega_1 w_0}{c_4}\left\{ -\zeta_3 Z+\frac{\pi^2}{6}\left[ \log(\mathcal{A})Z-\mathcal{I}(\omega_0,\omega_A; Z,\tau) \right] \right.  \addtocounter{equation}{1}\tag{\theequation}\label{eq:integralResultFinal}\\
    &\qquad \qquad\left. -\frac{1}{2}\log(\mathcal{A}) \left[ \mathcal{I}(\omega_0,\omega_A,\omega_C; Z,\tau)+\mathcal{I}(\omega_0,\omega_B,\omega_B; Z,\tau)-\mathcal{I}(\omega_0,\omega_A,\omega_A; Z,\tau) \right]  \right. \\
    &\qquad\qquad \left. +\frac{1}{2}\left[\mathcal{I}(\omega_0,\omega_A,\omega_C,\omega_C; Z,\tau)+\mathcal{I}(\omega_0,\omega_B,\omega_B,\omega_A; Z,\tau)-\mathcal{I}(\omega_0,\omega_A,\omega_A,\omega_A; Z,\tau)  \right] \right\} .
\end{align*}
Here we defined the algebraic function of the kinematic variables
\begin{equation}
     \mathcal{A}=\frac{1-y-z+\sqrt{\lambda(1,y,z)}}{2\sqrt{yz}}.
     \label{eq:algFctUChannel}
\end{equation}
Further we defined the (Chen) iterated integrals \cite{chenIteratedIntegrals}
\begin{equation}
    \mathcal{I}(\omega_1,\dots,\omega_n;x)=\int_{x_0}^x\mathrm{d}y_1\,f_1(y_1)\int_{x_0}^{y_1}\mathrm{d}y_2\,f_2(y_2)\dots \int_{x_0}^{y_{n-1}}\mathrm{d}y_n\,f_n(y_n),
\end{equation}
where the $f_i$ are defined to be the pull backs $f_i(y)\mathrm{d}y=\gamma^*\omega_i$ of the one-forms $\omega_i$ to the path $\gamma$ between the base point $x_0$ and the point $x$. The dependence on the path was left implicit above. In the Chen iterated integrals above we are taking the path from $x_0=(0,i\infty)$ along the straight line to $(0,\tau)$ and then along the straight line to the point $x=(Z,\tau)$. Since our result vanishes for $Z=0$ the first part does not contribute and the Chen iterated integrals reduce to eMPLs. \\ Expressed in terms of ordinary eMPLs the result reads
\begin{align*}
    &\frac{c_4}{\omega_1 w_0}I=-\zeta_3 Z +\frac{\pi^2}{6}\left\{ \log(\mathcal{A}) Z +\sum_{s=\pm}\left[\gamt{0 & 1}{0 & s\frac{1}{2}}{Z} - \gamt{0 & 1}{0 & 0}{Z}\right]\right\} \\
    &\quad +\frac{1}{2}\log(\mathcal{A})\sum_{s_1,s_2=\pm}\left[ -2\gamt{0 & 1 & 1}{0 & 0 & s_2\frac{1}{2}}{Z}+\gamt{0 & 1 & 1}{0 & 0 & s_2\frac{1+\tau}{2}}{Z} + \gamt{0 & 1 & 1}{0 & 0 & s_2\frac{\tau}{2}}{Z}\right. \\
    &\left. \qquad+ 2\gamt{0 & 1 & 1}{0 & s_1\frac{1}{2} & s_2\frac{1}{2}}{Z} - \gamt{0 & 1 & 1}{0 & s_1\frac{1}{2} & s_2\frac{1+\tau}{2}}{Z} -  \gamt{0 & 1 & 1}{0 & s_1\frac{1}{2} & s_2\frac{\tau}{2}}{Z} -  \gamt{0 & 1 & 1}{0 & s_1\frac{1+\tau}{2} & s_2\frac{1+\tau}{2}}{Z} \right. \\ 
    &\left. \qquad +\gamt{0 & 1 & 1}{0 & s_1\frac{1+\tau}{2} & s_2\frac{\tau}{2}}{Z} +  \gamt{0 & 1 & 1}{0 & s_1\frac{\tau}{2} & s_2\frac{1+\tau}{2}}{Z} -  \gamt{0 & 1 & 1}{0 & s_1\frac{\tau}{2} & s_2\frac{\tau}{2}}{Z}   \right] \\
    &\quad +\frac{1}{2}\sum_{s_1,s_2,s_3=\pm}\left[ 2 \gamt{0 & 1 & 1 & 1}{0 & 0 & 0 & s_3\frac{1}{2}}{Z}-\gamt{0 & 1 & 1 & 1}{0 & 0 & 0 & s_3\frac{1+\tau}{2}}{Z} - \gamt{0 & 1 & 1 & 1}{0 & 0 & 0 & s_3\frac{\tau}{2}}{Z} + 2\gamt{0 & 1 & 1 & 1}{0 & 0 & s_2\frac{1}{2} & 0}{Z} \right. \\
    &\left.\qquad - \gamt{0 & 1 & 1 & 1}{0 & 0 & s_2\frac{1}{2} & s_3\frac{1+\tau}{2}}{Z} - \gamt{0 & 1 & 1 & 1}{0 & 0 & s_2\frac{1}{2} & s_3\frac{\tau}{2}}{Z} - \gamt{0 & 1 & 1 & 1}{0 & 0 & s_2\frac{1+\tau}{2} & 0}{Z} - \gamt{0 & 1 & 1 & 1}{0 & 0 & s_2\frac{1+\tau}{2} & s_3\frac{1}{2}}{Z} \right. \\
    &\left. \qquad +\gamt{0 & 1 & 1 & 1}{0 & 0 & s_2\frac{1+\tau}{2} & s_3\frac{1+\tau}{2}}{Z} + \gamt{0 & 1 & 1 & 1}{0 & 0 & s_2\frac{1+\tau}{2} & s_3\frac{\tau}{2}}{Z} - \gamt{0 & 1 & 1 & 1}{0 & 0 & s_2\frac{\tau}{2} & 0}{Z} - \gamt{0 & 1 & 1 & 1}{0 & 0 & s_2\frac{\tau}{2} & s_3\frac{1}{2}}{Z} \right. \\
    &\left.\qquad +\gamt{0 & 1 & 1 & 1}{0 & 0 & s_2\frac{\tau}{2} & s_3\frac{1+\tau}{2}}{Z}  + \gamt{0 & 1 & 1 & 1}{0 & 0 & s_2\frac{\tau}{2} & s_3\frac{\tau}{2}}{Z} -2\gamt{0 & 1 & 1 & 1}{0 & s_1\frac{1}{2} & 0 & s_3\frac{1}{2}}{Z} + \gamt{0 & 1 & 1 & 1}{0 & s_1\frac{1}{2} & 0 & s_3\frac{1+\tau}{2}}{Z} \right. \\ 
    &\left.\qquad + \gamt{0 & 1 & 1 & 1}{0 & s_1\frac{1}{2} & 0 & s_3\frac{\tau}{2}}{Z} -2 \gamt{0 & 1 & 1 & 1}{0 & s_1\frac{1}{2} & s_2\frac{1}{2} & 0}{Z} +  \gamt{0 & 1 & 1 & 1}{0 & s_1\frac{1}{2} & s_2\frac{1}{2} & s_3\frac{1+\tau}{2}}{Z}  \right.  \\
    &\left.\qquad +\gamt{0 & 1 & 1 & 1}{0 & s_1\frac{1}{2} & s_2\frac{1+\tau}{2} & 0}{Z} + \gamt{0 & 1 & 1 & 1}{0 & s_1\frac{1}{2} & s_2\frac{1+\tau}{2} & s_3\frac{1}{2}}{Z} - \gamt{0 & 1 & 1 & 1}{0 & s_1\frac{1}{2} & s_2\frac{1+\tau}{2} & s_3\frac{1+\tau}{2}}{Z} \right. \\
    &\left. \qquad - \gamt{0 & 1 & 1 & 1}{0 & s_1\frac{1}{2} & s_2\frac{1+\tau}{2} & s_3\frac{\tau}{2}}{Z}  +\gamt{0 & 1 & 1 & 1}{0 & s_1\frac{1}{2} & s_2\frac{\tau}{2} & 0}{Z} +\gamt{0 & 1 & 1 & 1}{0 & s_1\frac{1}{2} & s_2\frac{\tau}{2} & s_3\frac{1}{2}}{Z} \right. \\
    &\left. \qquad - \gamt{0 & 1 & 1 & 1}{0 & s_1\frac{1}{2} & s_2\frac{\tau}{2} & s_3\frac{1+\tau}{2}}{Z}- \gamt{0 & 1 & 1 & 1}{0 & s_1\frac{1}{2} & s_2\frac{\tau}{2} & s_3\frac{\tau}{2}}{Z} + \gamt{0 & 1 & 1 & 1}{0 & s_1\frac{1+\tau}{2} & s_2\frac{1+\tau}{2} & 0}{Z} \right. \\
    &\left. \qquad - \gamt{0 & 1 & 1 & 1}{0 & s_1\frac{1+\tau}{2} & s_2\frac{1+\tau}{2} & s_3\frac{1}{2}}{Z}- \gamt{0 & 1 & 1 & 1}{0 & s_1\frac{1+\tau}{2} & s_2\frac{\tau}{2} & 0}{Z} + \gamt{0 & 1 & 1 & 1}{0 & s_1\frac{1+\tau}{2} & s_2\frac{\tau}{2} & s_3\frac{1}{2}}{Z} \right. \\
    &\left. \qquad - \gamt{0 & 1 & 1 & 1}{0 & s_1\frac{\tau}{2} & s_2\frac{1+\tau}{2} & 0}{Z} + \gamt{0 & 1 & 1 & 1}{0 & s_1\frac{\tau}{2} & s_2\frac{1+\tau}{2} & s_3\frac{1}{2}}{Z} + \gamt{0 & 1 & 1 & 1}{0 & s_1\frac{\tau}{2} & s_2\frac{\tau}{2} & 0}{Z}\right.\\
    &\left.\qquad - \gamt{0 & 1 & 1 & 1}{0 & s_1\frac{\tau}{2} & s_2\frac{\tau}{2} & s_3\frac{1}{2}}{Z} + \gamt{0 & 1 & 1 & 1}{0 & s_1\frac{1}{2} & s_2\frac{1}{2} & s_3\frac{\tau}{2}}{Z} \right].
    \addtocounter{equation}{1}\tag{\theequation}
    \label{eq:integralResultLong}
\end{align*}
Note that we can rewrite the prefactor of the result in terms of a modular form. Since this relation will play no further role in this paper, we will not review modular forms here but only give a brief introduction in appendix \ref{sec:modularForms} and also give further references there. Let us here just note that
\begin{equation}
    \frac{\omega_1 w_0}{c_4}= \frac{4\pi}{\sqrt{|\zeta-\zetab|}}\frac{\eta(2\tau)^4}{\eta(\tau)^2}\equiv \frac{4\pi}{\sqrt{|\zeta-\zetab|}}m(\tau),
    \label{eq:coefficientAsModularForm}
\end{equation}
where 
\begin{equation}
    \eta(\tau)=e^{\frac{i\pi\tau}{12}}\prod_{n=1}^\infty\left(1-e^{2\pi i n\tau}\right),
\end{equation}
is the Dedekind eta function and we made use of the $\zeta,\zetab$ variables introduced in eq.~\eqref{eq:crossRatios}. To find the $\eta$ quotient we performed a $q$-expansion of the left hand side making use of the known $q$-expansion of the modular lambda function \cite{OEIS} and identified the resulting sequence using the "On-Line Encyclopedia of Integer Sequences" \cite{OEIS} and ref.~\cite{etaQuotients}. The eta-quotient which we defined to be $m(\tau)$ now turns out to be a modular form, or more precisely an Eisenstein series, of weight 1 for the congruence subgroup $\Gamma(4)$. \\
Equations \eqref{eq:integralResultFinal} and \eqref{eq:integralResultLong} together with the relation \eqref{eq:coefficientAsModularForm} constitute the final form of the elliptic integral for the $u$-channel diagram. Note that the integral is not manifestly single-valued, i.e.~it is naturally expressed in terms of purely holomorphic periods \cite{zagierPeriods} and not in terms of single-valued periods \cite{brownSVPeriods} as one might have expected from the polylogarithmic parts of the amplitude.

\subsection{Result for the full one-loop amplitude}
\label{sec:fullAmplitude}
By combining the analytic result for the elliptic integral from the last subsection (and appendix \ref{sec:stChannel}) and the polylogarithmic contributions already computed in ref.~\cite{pierre} we can now present the full analytic result for the one-loop amplitude of a conformally coupled scalar in $\mathrm{AdS}_4$ dual to a primary operator of scaling dimension $\Delta=1$:
\begin{align*}
    &\left.(2\pi)^8x_{12}^2x_{34}^2\frac{4}{\lambda^2}\langle \phi_\Delta(x_1)\phi_\Delta(x_2)\phi_\Delta(x_3)\phi_\Delta(x_4) \rangle\right|_{\mathcal{O}(\lambda^2),\Delta=1}\\
    &~=\frac{\zeta\zetab}{\zeta-\zetab}\left\{-\frac{3}{\epsilon}\mathrm{BW}(\zeta,\zetab)-\frac{3}{2}f_1(\zeta,\zetab)+\frac{5}{2}\mathrm{BW}(\zeta,\zetab)\log(\zeta\zetab)-\frac{7}{2}\mathrm{BW}(\zeta,\zetab)\log((1-\zeta)(1-\zetab)\right\} \\
    &\qquad + \frac{4\pi\zeta\zetab}{\sqrt{|\zeta-\zetab|}}\sum_{i\in\{s,t,u\}}k_i m(\tau^{(i)})\left\{ -\zeta_3 Z^{(i)}+\frac{\pi^2}{6}\left[ \log(\mathcal{A}^{(i)})Z^{(i)}-\mathcal{I}(\omega_0,\omega_A; Z^{(i)},\tau^{(i)}) \right] \right. \\
    &\qquad \qquad\left. -\frac{1}{2}\log(\mathcal{A}^{(i)}) \left[ \mathcal{I}(\omega_0,\omega_A,\omega_C; Z^{(i)},\tau^{(i)})+\mathcal{I}(\omega_0,\omega_B,\omega_B; Z^{(i)},\tau^{(i)}) \right.  \right. \\
    &\qquad\qquad\qquad\qquad\qquad\quad \left. \left.-\mathcal{I}(\omega_0,\omega_A,\omega_A; Z^{(i)},\tau^{(i)}) -\mathcal{I}(\omega_0,\omega_A,\omega_A; Z^{(i)},\tau^{(i)}) \right] \right. \\
    &\qquad\qquad \left. +\frac{1}{2}\left[\mathcal{I}(\omega_0,\omega_A,\omega_C,\omega_C; Z^{(i)},\tau^{(i)})+\mathcal{I}(\omega_0,\omega_B,\omega_B,\omega_A; Z^{(i)},\tau^{(i)})  \right. \right. \\
    &\left. \left. \qquad\qquad\qquad\quad -\mathcal{I}(\omega_0,\omega_A,\omega_A,\omega_A; Z^{(i)},\tau^{(i)}) \right] \right\} +\mathcal{O}(\epsilon) .
    \addtocounter{equation}{1}\tag{\theequation}
    \label{eq:resultAmplitude}
\end{align*}
Here we defined
\begin{equation}
    k_s=\frac{1}{i\sqrt{\zeta\zetab}},\quad k_t=\frac{1}{i\sqrt{(1-\zeta)(1-\zetab)}},\quad k_u=1,
\end{equation}
and a pure, single-valued combination of multiple polylogarithms of weight 3 \cite{pierre}
\begin{align*}
    f_1(\zeta,\zetab)&=\log(\zeta\zetab)\left[\mathrm{Li}_{1,1}\left( \zetab,\frac{\zeta}{\zetab}\right) -\mathrm{Li}_{1,1}\left( \zeta,\frac{\zetab}{\zeta}\right)+\mathrm{Li}_1(\zeta)\mathrm{Li}_1\left(\frac{\zetab}{\zeta}\right)-\mathrm{Li}_1(\zetab)\mathrm{Li}_1\left(\frac{\zeta}{\zetab}\right)\right] \\
    &\quad +2\mathrm{Li}_{2,1}\left( \zeta,\frac{\zetab}{\zeta}\right)-2\mathrm{Li}_{2,1}\left( \zetab,\frac{\zeta}{\zetab}\right)+\mathrm{Li}_{1,2}\left( \zeta,\frac{\zetab}{\zeta}\right)-\mathrm{Li}_{1,2}\left( \zetab,\frac{\zeta}{\zetab}\right) \\
    &\quad +2\mathrm{Li}_1\left(\frac{\zeta}{\zetab}\right)\mathrm{Li}_2(\zetab)-2\mathrm{Li}_1\left(\frac{\zetab}{\zeta}\right)\mathrm{Li}_2(\zeta)+\mathrm{Li}_1(\zetab)\mathrm{Li}_2\left(\frac{\zeta}{\zetab}\right)-\mathrm{Li}_1(\zeta)\mathrm{Li}_2\left(\frac{\zetab}{\zeta}\right) \\
    &\quad +\mathrm{Li}_3(\zeta)-\mathrm{Li}_3(\zetab)+\mathrm{Li}_{2,1}(1,\zeta)-\mathrm{Li}_{2,1}(1,\zetab).
    \addtocounter{equation}{1}\tag{\theequation}
    \label{eq:f1Function}
\end{align*}
Also recall that $\mathrm{BW}(\zeta,\zetab)$ refers to the Bloch-Wigner dilogarithm \eqref{eq:blochWigner} and $m(\tau)$ to the modular form \eqref{eq:coefficientAsModularForm}. By expressing $x,y,z$ as conformal cross ratios $u,v$ (following the prescription \eqref{eq:integralChannels}) in eqs.~\eqref{eq:algFctSChannel},~\eqref{eq:algFctTChannel} and \eqref{eq:algFctUChannel} we can write
\begin{equation}
\begin{split}
    \mathcal{A}^{(s)}&=\frac{u-v-1-\sqrt{\lambda(1,u,v)}}{2\sqrt{v}}, \\
    \mathcal{A}^{(t)}&=\frac{v-u-1-\sqrt{\lambda(1,u,v)}}{2\sqrt{u}}, \\
    \mathcal{A}^{(u)}&=\frac{1-u-v+\sqrt{\lambda(1,u,v)}}{2\sqrt{uv}}.
\end{split}
\end{equation}
Note that the result contains three different pairs of variables $(Z,\tau)$ indexed by the three channels because they are computed using different elliptic curves, see appendix \ref{sec:stChannel}. The $j$-invariants of the elliptic curves corresponding to the $s$-,$t$- and $u$-channel integrals respectively, are given by
\begin{equation}
\begin{split}
    j_s&=\frac{256(\lambda(1,u,v)+v)^3}{v^2\lambda(1,u,v)}, \\
    j_t&=\frac{256(\lambda(1,u,v)+u)^3}{u^2\lambda(1,u,v)}, \\
    j_u&=\frac{256(\lambda(1,u,v)+uv)^3}{u^2v^2\lambda(1,u,v)}.
    \label{eq:jInvariants}
\end{split}
\end{equation}
Indeed, for a generic kinematic point the three elliptic curves are not isomorphic.\\
Equation \eqref{eq:resultAmplitude} adds to the very short list of known one-loop curved-space amplitudes in position space. Also in contrast to the majority of loop-level results which build upon analytic conformal bootstrap methods \cite{analyticCFTBootstrapReview}, this amplitude was computed purely from the bulk side. \\
A couple of comments are in order. First note that both the polylogarithmic and the elliptic parts are pure combinations of (e)MPLs in the sense of Refs.~\cite{nimaPure,hennCanonicalBasis,claudeEllipticPure}. They are also both of uniform transcendentality, all terms having weight 3 (when one assigns $\epsilon$ weight -1, as is customary). This is not the case for the amplitude with $\Delta=2$ where the maximum transcendental weight is 3 but also terms of lower transcendental weight appear \cite{pierre}.\\
There is an important qualitative difference between the polylogarithmic and the elliptic parts of the $\Delta=1$ amplitude. While the polylogarithmic part is naturally expressed in terms of single-valued MPLs, which are examples of single-valued periods, and hence is manifestly single-valued, this is not true for the elliptic part. Here only holomorphic periods occur spoiling the manifest single-valuedness of the amplitude. This explicitly shows that manifest single-valuedness in the polylogarithmic sector does not have to be an indication for the same property in the elliptic sector. Again this is different for the $\Delta=2$ amplitude which is purely polylogarithmic and manifestly single-valued \cite{pierre}.\\
One might be worried about the seemingly divergent limit $\zetab\rightarrow \zeta$, where the four boundary points approach a line, i.e.~a one-dimensional configuration. It turns out, however, that the apparent singularity cancels in the limit. This is easy to see in the polylogarithmic part but not immediately obvious in the elliptic sector.\\
For simplicity let us again focus on the $u$-channel contribution. The apparent singularity is sitting in the overall prefactor but turns out to be cancelled by a zero in the modular form $m(\tau)$. This is most easily seen by just studying the fraction $\omega_1 w_0/c_4$ in the limit. A short computation shows that it is indeed finite
\begin{equation}
    \frac{\omega_1 w_0}{c_4}\rightarrow\frac{\pi}{\sqrt{\zeta(1-\zeta)}}.
\end{equation}
The finiteness of the prefactor in the elliptic contribution should be contrasted with the divergent prefactor in the polylogarithmic contribution which only gets cancelled by a zero in the polylogarithms. This
is connected to the manifest single-valuedness of the polylogarithmic contribution leading to a zero in the $\zetab\rightarrow \zeta$ limit which is cancelled by the pole in the prefactor. Since the elliptic sector does not have this zero in the eMPLs there is no pole in the prefactor.\\
Further note that also the eMPLs simplify in this limit. To understand this note that a straightforward computation shows that the limit in the variables on the elliptic curve is given by the modular parameter $\tau$ approaching the cusp at infinity and the puncture $Z$ approaching a transcendental function of the cross ratio $\zeta$
\begin{equation}
    \tau\rightarrow i\infty,\quad Z\rightarrow -\frac{1}{\pi}\arctan\left(\frac{\sqrt{1-4\zeta(1-\zeta)}-1}{2\sqrt{\zeta(1-\zeta)}}\right).
\end{equation}
Hence to compute the value of the eMPLs in the limit we need to replace all of the eMPLs with their cusp value, which can be explicitly computed in terms of ordinary MPLs using the $q$-expansion of the integration kernels \cite{claudeEllipticSymbol} and in particular is finite\footnote{An explicit computation shows that harmonic polylogarithms \cite{remiddiHPLs} are actually enough to express the result.}. Hence we conclude that in the limit of a one dimensional configuration of boundary points the result is finite and reduces to a combination of MPLs.\\
As we made no use of any CFT information in our computation it is an interesting question to ask what the amplitude now implies for the boundary CFT. As is well known, the interaction in the bulk leads to quantum corrections to the scaling dimensions of the operators on the boundary, i.e.~anomalous dimensions. Using the operator product expansion (OPE) of the correlation function we just computed it is now possible to extract the first order anomalous dimension of the operator dual to the scalar field $\phi$. This was fully carried out in ref.~\cite{pierre}, as the extraction of anomalous dimension does not need the full evaluation of the elliptic integral, and we refer the reader to this reference for further details. \\
One might however ask how these results can be obtained from our full analytic results for the correlation function. As discussed in ref.~\cite{pierre}, the anomalous dimension can be read off from the coefficient of $\log(u)$ in the OPE limit $x_1\rightarrow x_2,~x_3\rightarrow x_4$ of the correlation function. We will not fully perform this computation as the results were already obtained in ref.~\cite{pierre}, we will however quickly explain how one could reproduce the anomalous dimension here, i.e.~how the limit acts on our result.\\
We will again restrict to the $u$-channel, hence we are interested in the $\log(y)$ expansion. Note that the $y\rightarrow 0$ limit is the degeneration limit of the elliptic curve where the two positive and the two negative roots coincide and the holomorphic differential degenerates to a sum of $\mathrm{d}\!\log$ forms. Hence the integrations in the definition of the periods can now introduce logarithmic end-point singularities which are precisely the logarithms we are after. More concretely we find the following leading behaviour in the $y\rightarrow 0$ limit:
\begin{equation}
    \tau\sim\frac{-i\pi}{\log y},~Z\sim\frac{1}{2\log y}\log\left(\frac{1-\sqrt{1-z}}{1+\sqrt{1-z}} \right),~ \log\mathcal{A}\sim -\frac{1}{2}\log y,~ \frac{\omega_1 w_0}{c_4}\sim -\frac{1}{(1-z)^{3/2}}\log y .
\end{equation}
Hence we see that $\tau$ and $Z$ vanish in the limit while the prefactor (leading singularity) and the logarithms of the result diverge. \\
To find the behaviour of the eMPLs in the limit we can use a modular transformation $\tau= -1/\tau'$ and then study the limit $\tau'\rightarrow i\infty$. The expansion in this limit just corresponds to the well-known $q$-expansion. One can use the modular properties of the kernels and their $q$-expansions as given in ref.~\cite{claudeEllipticSymbol} to find the behaviour of the eMPLs in the limit. Let us illustrate this with a simple example:
\begin{align*}
    \gamt{1}{\frac{1}{2}}{Z;-\frac{1}{\tau'}}&=\int_0^Z\mathrm{d}z\,g^{(1)}(z-1/2,-1/\tau') \\
    &=\int_0^Z\mathrm{d}z\,\left[-\tau' g^{(1)}(-\tau' z+\tau'/2,\tau')+2\pi i \tau'(z-1/2) \right] \\
    &=i\pi\tau' Z^2-\log\left(1-e^{-i\pi\tau'}\right)+\log\left(1-e^{-i\pi\tau'+2\pi i\tau' Z} \right)+\mathcal{O}\left(e^{i\pi\tau'}\right) \\
    &\sim i\pi\tau' Z^2=-\frac{i\pi Z^2}{\tau}. \addtocounter{equation}{1}\tag{\theequation}
\end{align*}
One could work out the other eMPLs in a similar way and hence compute the anomalous dimension of the primary operator dual to the bulk scalar field.

\section{The symbol of the elliptic integral}
\label{sec:symbol}
In the last section we have found an explicit expression for the elliptic integral $I(1,y,z)$ in terms of eMPLs. This now allows us to straightforwardly compute its symbol as reviewed in section \ref{ssec:symbolReview}.\\
After computing the symbol and projecting onto the highest length part we can immediately see a big simplification just by making use of the trivial relations of the symbol letters, such that the only letters appearing are of the form $\Omega^{(n)}$ with $n\leq 2$. Inspired by the simplification of the result of the integral in the previous section after identifying certain combinations of kernels we analogously define the following combinations of symbol letters
\begin{align*}
        \Omega^{(n)}_A(z,\tau)&=\Omega^{(n)}(z,\tau)-\Omega^{(n)}(z-1/2,\tau), \\
        \Omega^{(n)}_B(z,\tau)&=\Omega^{(n)}(z-\tau/2,\tau)-\Omega^{(n)}(z-1/2-\tau/2,\tau), \addtocounter{equation}{1}\tag{\theequation} \\
        \Omega^{(n)}_C(z,\tau)&=\Omega^{(n)}(z,\tau)+\Omega^{(n)}(z-1/2,\tau)-\Omega^{(n)}(z-\tau/2,\tau)-\Omega^{(n)}(z-1/2-\tau/2,\tau).
\end{align*}
Using the identity connecting certain combinations of letters to a $\dd\!\log$ form (and a limiting case) we can express all combinations of letters with $n=1$ appearing in the symbol as logarithms. We can then write the symbol in the following form\footnote{I am grateful to Matthias Wilhelm and Chi Zhang for pointing this out.}
\begin{equation}
    \mathcal{S}\left(16\pi i\frac{\omega_1 w_0}{c_4} I(1,y,z)\right)=\sum_{i=1}^2\mathcal{S}(f_i)\otimes\left[\log a_i\otimes\Omega^{(0)}(Z,\tau)+\Theta_i\otimes (2\pi i\tau)\right],
    \label{eq:symbol}
\end{equation}
where 
\begin{align}
    \mathcal{S}(f_1)&=\log(y)\otimes\log(\mathcal{A}\mathcal{B})+\log(z)\otimes\log(\mathcal{A}/\mathcal{B}), \\
    \mathcal{S}(f_2)&=\log(y)\otimes\log(z)+\log(z)\otimes\log(y),
\end{align}
and
\begin{alignat}{3}
    a_1&=\mathcal{B}^2, \quad &&\Theta_1=4\Omega_B^{(2)}(Z,\tau)-\log\mathcal{B}, \\
    a_2&=y/z, \quad &&\Theta_2=4\Omega_A^{(2)}.
\end{alignat}
Here we defined the algebraic function
\begin{equation}
    \mathcal{B}=\frac{y+z-(y-z)\left(y-z-\sqrt{\lambda(1,y,z)}\right)}{2\sqrt{y z}},
\end{equation}
also recall that $\mathcal{A}$ was defined in eq.~\eqref{eq:algFctUChannel}. \\
There are a few comments in order. First note that all first entries of the symbol are given by logarithms of the conformal cross ratios, which is precisely the physical first entry condition \cite{firstEntryCondition}. Also the second entries are purely logarithmic but can be complicated algebraic functions of the cross ratios.
The final entries are only given by the basic elliptic symbol letters
\begin{equation}
    \Omega^{(0)}(Z,\tau)=2\pi i Z,\quad\Omega^{(-1)}(0,\tau)=-2\pi i\tau.
\end{equation}
These features where also observed in the elliptic flat-space 10-point double-box integral in ref.~\cite{wilhelmDoubleBoxSymbol}.\\
When looking at the third entries, we observe both simple logarithmic letters, accompanied by the $\dd Z$ form in the last entry, as well as more complicated elliptic letters, accompanied by the $\dd\tau$ form in the final entry. These very different looking logarithmic and elliptic third entries are however very closely related, which is made manifest by the symbol prime \cite{wilhelmSymbolPrime}
\begin{equation}
    \mathcal{S}'(\Theta_i)=\Omega^{(0)}(Z,\tau)\otimes' \log(a_i).
\end{equation}
This relation further makes the form \eqref{eq:symbol} of the symbol manifestly doubly periodic and also manifests some of the integrability properties, see ref.~\cite{wilhelmSymbolPrime} for a more elaborate discussion on these points. \\
This relationship between logarithmic and elliptic third entries was first observed in the context of flat-space amplitudes for the unequal mass sunrise and 10-point double-box integrals \cite{wilhelmSymbolPrime}. The structure and in particular the relation between the elliptic and logarithmic third entries was then used to bootstrap the symbol of the 12-point double-box integral in ref.~\cite{wilhelmSymbolBootstrap}.

\section{Conclusions and future directions}
\label{sec:conclusions}
The main result of this paper is the computation of the elliptic integral contributing to the finite part of the four-point amplitude of a conformally coupled scalar in Euclidean $\mathrm{AdS}_4$ dual to a primary operator with scaling dimension $\Delta=1$ in the dual CFT. This computation allowed us to present the full analytic result for the amplitude, serving as one of the rare examples of an explicit result for a curved space amplitude beyond tree level computed directly in position space. \\
Furthermore to the best of our knowledge this is the first instance of a curved-space amplitude
evaluated in terms of elliptic multiple polylogarithms which play an increasingly important role in flat-space computations. Another interesting observation is that the computed integral spoils the manifest single-valuedness of the amplitude since only a holomorphic variable appears in our result. \\
We furthermore computed the symbol of the integral, finding similar structures as observed for the symbol of the flat-space double-box integral. In particular the last entry only consists of basic elliptic symbol letters, and the first entry only contains logarithms of conformal cross ratios in accordance with the physical first entry condition. Furthermore we observe that the symbol prime relates the complicated elliptic third entries to the simple logarithmic ones, which was first observed in ref.~\cite{wilhelmSymbolPrime}.\\
Furthermore we note that the correspondence between Witten diagrams and flat-space Feynman integrals that builds the basis for the computation of the amplitude in this paper seems to be very special, i.e.~the setup of the theory is quite specific. One might hence wonder if there is a more general connection, which would allow to import methods from flat-space multi-loop Feynman integral computations to compute more general position space amplitudes in AdS. \\
Let us point out that the key player of the correspondence of ref.~\cite{pierre} is the conformal coupling and it is quite easy to vary other parameters such as the boundary dimension $d$ (and hence the scaling dimensions $\Delta$) or the valency of the vertex $n$. To see this note that for a conformally coupled scalar in $d$ dimensions the allowed scaling dimensions are $\Delta_\pm=(d\pm 1)/2$. If one plugs these values into the general form of the bulk-to-bulk propagator one easily finds that it again takes the form of a flat-space propagator (with higher propagator powers).\\
A bit of care still has to be taken concerning the parity of the integrand in the radial coordinates $z$, which we needed to be even to extend the integration domain. To study this note that a vertex with valency $n$ will contribute a factor (for simplicity assuming all scaling dimensions at the vertex to be equal) $z^{n\Delta_\pm-(d+1)}$. From this it easily follows that for all odd $d$ and arbitrary $n$ at least one of the two scaling dimensions $\Delta_\pm$ will lead to an even integrand, hence allowing the use of flat-space methods. Even $d$ are more tricky since they lead to non-integer scaling dimensions which do not allow a naive extension of the integration domain. \\
Hence there is a wider class of possible diagrams with different boundary and scaling dimensions as well as different vertex valencies which can be attacked with flat-space methods such as the ones reviewed and used in this paper. It would certainly be an interesting question how far one can push this for conformally coupled scalars. In particular it would be interesting to study higher point diagrams which are, however, technically quite challenging due to the high number of variables.\\
The more interesting question however would be if one can also find a connection between Witten diagrams and flat-space Feynman integrals beyond conformally coupled scalars. This has been fully understood for contact diagrams \cite{zhouYangianCorrelators} but connections between exchange or loop diagrams are, to the best of our knowledge, restricted to conformal coupling so far.

\acknowledgments

I would like to thank Claude Duhr for proposing this project and giving many valuable suggestions. Furthermore, I would like to thank Florian Loebbert for discussions and comments on the draft as well as Ekta Chaubey and Yu Jiao Zhu for discussions. Also I would like to thank Matthias Wilhelm and Chi Zhang for a very helpful comment and clarification concerning the structure of the elliptic symbol and the role of the symbol prime. \\
The work of SFS is supported by funds of the Klaus Tschira Foundation gGmbH. 

\appendix
\section{The integral in the $s$- and $t$-channel}
\label{sec:stChannel}
In section \ref{sec:ellipticIntegral} we computed the elliptic integral $I(x,y,z)$ in the kinematic region 
\begin{equation}
    \sqrt{y}+\sqrt{z}<\sqrt{x},
\end{equation}
for $x=1$, which in particular includes the correct parameters for the $u$-channel contribution to the amplitude in the kinematic region $\sqrt{u}+\sqrt{v}<1$. In this appendix we want to sketch the computation for the remaining $s$- and $t$-channel contributions in the same kinematic region.\\
Since the integral is symmetric in $y,z$ it makes no difference if we choose $z=1$ and $\sqrt{x}+\sqrt{y}<1$ corresponding to the above kinematic region in the $s$-channel or with $y$ and $z$ interchanged corresponding to the $t$-channel. For definiteness let us thus set $z=1$ but the result will also be valid in the $t$-channel (with the role of $u,v$ exchanged).\\
The main difference in the computation is that the branch points of the elliptic curve in this case are not all real but all imaginary. We can however simply perform a coordinate transformation $\sigma=-i\sigma'$ which leads to a new form of the elliptic curve
\begin{equation}
    (w^{(s)})^2=P_4^{(s)}(\sigma),
\end{equation}
with $P_4^{(s)}(\sigma)=\left. P_4(-i\sigma)\right|_{z=1}$. In these coordinates the branch points are now all real again. The steps to compute the integral are now exactly the same as before and the result also takes the same form
\begin{align*}
    I&=\frac{\omega_1^{(s)} w_0^{(s)}}{ic_4^{(s)} x}\left\{-\zeta_3 Z^{(s)}+\frac{\pi^2}{6}\left[ \log(\mathcal{A}^{(s)})Z^{(s)}-\mathcal{I}(\omega_0,\omega_A; Z^{(s)},\tau^{(s)}) \right] \right. \\
    &\qquad\qquad\qquad \left. -\frac{1}{2}\log(\mathcal{A}^{(s)}) \left[ \mathcal{I}(\omega_0,\omega_A,\omega_C; Z^{(s)},\tau^{(s)})+\mathcal{I}(\omega_0,\omega_B,\omega_B; Z^{(s)},\tau^{(s)}) \right.\right.\\
    &\qquad\qquad\qquad\qquad\qquad\qquad\quad\left.\left.-\mathcal{I}(\omega_0,\omega_A,\omega_A; Z^{(s)},\tau^{(s)}) \right]  \right. \\
    &\qquad\qquad\qquad \left. +\frac{1}{2}\left[\mathcal{I}(\omega_0,\omega_A,\omega_C,\omega_C; Z^{(s)},\tau^{(s)})+\mathcal{I}(\omega_0,\omega_B,\omega_B,\omega_A; Z^{(s)},\tau^{(s)})\right.\right. \\
    &\left.\left.\qquad\qquad\qquad\qquad\quad -\mathcal{I}(\omega_0,\omega_A,\omega_A,\omega_A; Z^{(s)},\tau^{(s)})  \right]\right\}, \addtocounter{equation}{1}\tag{\theequation}
\end{align*}
with
\begin{equation}
    Z^{(s)}=\frac{c_4^{(s)}}{\omega_1^{(s)}}\int_0^i\frac{\mathrm{d}\sigma}{w^{(s)}},
\end{equation}
and
\begin{equation}
     \mathcal{A}^{(s)}=\frac{x-y-1-\sqrt{\lambda(x,y,1)}}{2\sqrt{y}}.
     \label{eq:algFctSChannel}
\end{equation}
In the $t$-channel one has 
\begin{equation}
    \mathcal{A}^{(t)}=\frac{x-1-z-\sqrt{\lambda(x,1,z)}}{2\sqrt{z}}.
     \label{eq:algFctTChannel}
\end{equation}
Furthermore we find
\begin{align}
    \frac{\omega_1^{(s)} w_0^{(s)}}{i c_4^{(s)} x}&=\frac{4\pi}{\sqrt{|\zeta-\zetab|}} \frac{1}{i\sqrt{\zeta\zetab}}m(\tau^{(s)}), \\
    \frac{\omega_1^{(t)} w_0^{(t)}}{i c_4^{(t)} x}&=\frac{4\pi}{\sqrt{|\zeta-\zetab|}} \frac{1}{i\sqrt{(1-\zeta)(1-\zetab)}}m(\tau^{(t)}),
\end{align}
in $s$- and $t$-channel respectively. All quantities are computed on the transformed elliptic curve with real branch points, we could however also transform back to the original curve.\\
Note that the three elliptic curves in the $s$-,$t$-,and $u$-channel are non-isomorphic for generic $u,v$, as can be seen from their respective $j$-invariants given in eq.~\eqref{eq:jInvariants}. Hence we have three different sets of $\tau$ and $Z$ parameters. They have to be related since the physical problem only has two degrees of freedom but it is not possible to write down an explicit relation between them. Implicit relations can be obtained for example by using the explicit form of the $j$-invariants as rational functions of the conformal cross ratios to derive relations between the three $j$ invariants which can then be read as implicit relations between the $\tau$ parameters. Similarly one can compute $\kappa(Z)$ for the three $Z$s in terms of the cross ratios and derive implicit relations for the $Z$ variables.

\section{Technical details of the calculation}
\label{sec:detailsOfCalculation}
In section \ref{sec:ellipticIntegral} when computing the elliptic integral we stated that to get to the final form \eqref{eq:integralResultFinal} of the integral we need to rewrite certain combinations of length one eMPLs as logarithms. The only combination appearing at length one is 
\begin{equation}
    \mathcal{I}(\omega_C;Z,\tau)=\int_0^Z\mathrm{d}z\,\left( g_\pm^{(1)}(z,0,\tau)+ g_\pm^{(1)}(z,1/2,\tau)-g_\pm^{(1)}(z,\tau/2,\tau)-g_\pm^{(1)}(z,1/2+\tau/2,\tau)\right).
\end{equation}
The difference of two terms here can always be rewritten as a logarithm. If no regularization is involved this is straightforward. For example we find
\begin{equation}
    \int_0^Z\mathrm{d}z\,\left( g_\pm^{(1)}(z,1/2,\tau)-g_\pm^{(1)}(z,1/2+\tau/2,\tau)\right)=\log\left( \frac{\kappa(Z)-a_4}{2a_4}\frac{a_3+a_4}{\kappa(Z)-a_3}\right),
\end{equation}
where $\kappa(Z)$ was given in eq.~\eqref{eq:kappaOfZ}. However when the regularization is involved there are some technical complications that will be described in more detail in the rest of this appendix. \\
First note that at the start of the computation we had no need to regularize as everything was manifestly finite. Only after shuffling all the $\left( \begin{smallmatrix} 0 \\ 0 \end{smallmatrix}\right)$ entries to the left we encounter divergent eMPLs, their regularization however can be chosen arbitrarily because they would of course again all drop out by using the shuffle algebra. When we now want to rewrite certain (regularized) combinations of eMPLs as logarithms we will have to choose a concrete regularization and the choice will play a role for the result. The reason for choosing a "non-standard" regularization as indicated in section \ref{sec:elliptic} will become clear in the following. \\
We need to compute one logarithm involving a regularized eMPL and we can choose that one to be given by the combination
\begin{equation}
    L_\mathrm{reg}=\int_0^Z\mathrm{d}z\,\left( g_\pm^{(1)}(z,0,\tau)-g_\pm^{(1)}(z,1/2,\tau) \right).
\end{equation}
Let us first compute $L_\mathrm{reg}$ by regularizing the first term as
\begin{equation}
    \mathrm{R}\left[ \int_{\epsilon/\omega_1}^Z\mathrm{d}z\,g_\pm^{(1)}(z,0,\tau) \right],
\end{equation}
where $\mathrm{R}$ is an operator picking out the constant term in a $\log\epsilon$ expansion. One finds
\begin{equation}
    \mathrm{R}\left[\int_{\epsilon/\omega_1}^Z\mathrm{d}z\,\left( g_\pm^{(1)}(z,0,\tau)-g_\pm^{(1)}(z,1/2,\tau) \right) \right]=\log\left( \frac{(\kappa(Z)-a_1)(a_3+a_4)}{2a_4(a_3-a_4)}\frac{-2a_4}{\kappa(Z)-a_4} \right).
\end{equation}
This now differs from the result in the "usual" regularization as for example chosen in ref.~\cite{claudeEllipticPure} by some $C(\tau)$ which is constant in $Z$. This is now computed in the usual regularization
\begin{equation}
    C(\tau)=\int_0^{1/4}\mathrm{d}z\,\left( g_\pm^{(1)}(z,0,\tau)-g_\pm^{(1)}(z,1/2,\tau) \right)-\log\left( \frac{(\kappa(1/4)-a_1)(a_3+a_4)}{2a_4(a_3-a_4)}\frac{-2a_4}{\kappa(1/4)-a_4} \right),
\end{equation}
where we conveniently put $Z=1/4$ since the full expression is independent of $Z$. The second term is easily evaluated using $\kappa(1/4)=\infty$ yielding
\begin{equation}
    \log\left( \frac{a_4+a_3}{a_4-a_3} \right).
\end{equation}
The integral we are now left to evaluate
\begin{equation}
    F(\tau)=2\int_0^{1/4}\mathrm{d}z\,\left( g^{(1)}(z,\tau)-g^{(1)}(z-1/2,\tau) \right),
\end{equation}
is a bit tricky. By differentiating and integrating back up one can write it as
\begin{equation}
    F(\tau)=2\int_{i\infty}^\tau\frac{\mathrm{d}\tau'}{2\pi i}\left[ g^{(2)}(1/2,\tau')-g^{(2)}(0,\tau')\right]-i\pi+2\log(2) .
\end{equation}
Here the cusp value can be computed from the $q$-expansions of the $g$ kernels as can be found in ref.~\cite{claudeEllipticSymbol} for example and the derivative can be evaluated using the mixed-heat equation property of the kernels \cite{claudeEllipticSymbol}
\begin{equation}
    \frac{\partial}{\partial \tau}g^{(n)}(z,\tau)=\frac{n}{2\pi i}\frac{\partial}{\partial z}g^{(n+1)}(z,\tau).
\end{equation}
The final integral can be computed explicitly and finally yields \cite{claudeEisensteinTools}
\begin{equation}
    F(\tau)=-i\pi-2\log\left(\frac{K(\lambda(\tau))}{\pi}\right)-\frac{1}{2}\log(1-\lambda(\tau)).
\end{equation}
Here $K$ is the complete elliptic integral of the first kind and $\lambda$ is the modular lambda function. In terms of branch points and periods these are given by
\begin{align}
    \lambda(\tau)&=\frac{(a_1-a_4)(a_2-a_3)}{(a_1-a_3)(a_2-a_4)}, \\
    K(\lambda(\tau))&=\omega_1/2,
\end{align}
so that by putting everything together we finally get
\begin{equation}
    C(\tau)=2\log\left( \frac{-2\pi i}{\omega_1} \right).
\end{equation}
If we now absorb this into the regularization we get the clean relation
\begin{equation}
    \int_{0}^Z\mathrm{d}z\,\left( g_\pm^{(1)}(z,0,\tau)-g_\pm^{(1)}(z,1/2,\tau) \right) =\log\left( \frac{(\kappa(Z)-a_1)(a_3+a_4)}{2a_4(a_3-a_4)}\frac{-2a_4}{\kappa(Z)-a_4} \right).
\end{equation}
where now the divergent eMPLs are regularized with the non-standard regularization as given in section \ref{sec:elliptic}. Combining this with the other logarithms arising from eMPLs that do not require regularization then yields the logarithm that shows up in the final result \eqref{eq:integralResultFinal}.

\section{Brief introduction to modular forms}
\label{sec:modularForms}
In this appendix we introduce some basics of modular forms necessary to understand the relation \eqref{eq:coefficientAsModularForm}. For a more rigorous treatment consider Refs.~\cite{modularFormsBook,zagierModularForms}, also see Refs.~\cite{claudeEllipticSymbol,weinzierlModularForms,claudeModularBanana} for more elaborate introductions in the physics literature. \\
First we fix a so-called congruence subgroup $\Gamma$ of the modular group $\mathrm{SL}(2,\mathbb{Z})$. This is a subgroup that contains a principal congruence subgroup
\begin{equation}
    \Gamma(N)=\left\{ M\in \mathrm{SL}(2,\mathbb{Z}),~A\equiv \begin{pmatrix}
        1 & 0 \\ 0 & 1   \end{pmatrix} \mathrm{mod}~N \right\} \trianglelefteq \mathrm{SL}(2,\mathbb{Z}).
\end{equation}
The action of the congruence subgroup $\Gamma$ on the extended upper-half plane $\mathbb{H}\cup\mathbb{Q}\cup\{i\infty\}$ partitions $\mathbb{Q}\cup\{i\infty\}$ into distinct orbits, to which we will refer (often in the form of a representative) as the cusps of $\Gamma$. \\
We can now define a modular form of weight $k$ for the congruence subgroup $\Gamma\leqslant\mathrm{SL}(2,\mathbb{Z})$ to be a holomorphic function $f:\mathbb{H}\rightarrow \mathbb{C}$ which transforms covariantly under the action of $\Gamma$
\begin{equation}
    f\left( \frac{a\tau+b}{c\tau+d}\right)=(c\tau+d)^k f(\tau) \text{ for all } \begin{pmatrix}
          a & b \\
          c & d
    \end{pmatrix}\in\Gamma,
\end{equation}
and is holomorphic at all cusps of $\Gamma$. We will denote the space of modular forms of weight $k$ for the congruence subgroup $\Gamma$ by $\mathcal{M}_k(\Gamma)$.\\
Let us briefly comment on the last condition. Every modular form for $\Gamma(N)$ admits a Fourier expansion
\begin{equation}
    f(\tau)=\sum_{n=-\infty}^\infty a_n q_N^n,~\text{with } q_N\equiv q^{1/N}\equiv e^{2\pi i\tau/N}.
\end{equation}
The function $f$ is now said to be holomorphic at the cusp $i\infty$ if there are no negative powers of $q_N$ in this expansion. The holomorphicity at the other cusps can be understood by mapping the cusps to $i\infty$, i.e.~$f$ is holomorphic at the cusp $r\in\mathbb{Q}$ if
\begin{equation}
    \left[f|_k\gamma \right](\tau)\equiv(c\tau+d)^{-k}f\left(\frac{a\tau+b}{c\tau+d}\right),\quad \gamma=\begin{pmatrix}
        a & b \\ c & d
    \end{pmatrix}\in\mathrm{SL}(2,\mathbb{Z}),
\end{equation}
is holomorphic at $i\infty$, where $\gamma\in\mathrm{SL}(2,\mathbb{Z})$ is chosen such that $r=\gamma\cdot i\infty\equiv a/c$. \\
If a modular form $f\in\mathcal{M}_k(\Gamma)$ vanishes at all of the cusps of $\Gamma$ we refer to it as a cusp form. Let $\mathcal{S}_k(\Gamma)$ be the space of all cusp forms of weight $k$ for the congruence subgroup $\Gamma$. Then the space of modular forms decomposes into a direct sum of the space of cusp forms and the so-called Eisenstein subspace $\mathcal{E}_k(\Gamma)$
\begin{equation}
    \mathcal{M}_k(\Gamma)=\mathcal{S}_k(\Gamma)\oplus\mathcal{E}_k(\Gamma),
\end{equation}
The elements of $\mathcal{E}_k(\Gamma)$ are referred to as Eisenstein series of weight $k$ for the congruence subgroup $\Gamma$.




\bibliographystyle{JHEP}
\bibliography{biblio.bib}

\end{document}